\documentclass[11pt,a4paper]{amsart}

\usepackage{amsmath,amssymb,amsthm}
\usepackage{latexsym}

\newcommand{\HH}{{\mathcal H}}
\newcommand{\LpH}{{L^{p}_{\HH}}}
\newcommand{\LuH}{{L^{1}_{\HH}}}
\newcommand{\LiH}{{L^{\infty}_{\HH}}}

\newcommand{\uno}{{\mathbf{1}}}

\newcommand{\R}{\mathbb R}

\newcommand{\xx}{\langle x\rangle}

\newcommand{\supp}{\mathop{\mathrm{supp}}}

\numberwithin{equation}{section}

\newtheorem{thm}{Theorem}[section]
\newtheorem{cor}{Corollary}[section]
\newtheorem{lem}{Lemma}[section]
\newtheorem{prop}{Proposition}[section]

\theoremstyle{remark}

\theoremstyle{definition}

\theoremstyle{remark}
\newtheorem{remark}{Remark}[section]

\theoremstyle{remark}
\newtheorem*{ack}{Acknowledgments}

\begin{document}
\baselineskip15pt

\title[$L^{p}$-boundedness of the wave operator]
{$L^p$-boundedness of the wave operator for the
one dimensional Schr\"odinger operator}

\author{Piero D'Ancona, Luca Fanelli}

\thanks{Piero D'Ancona, Dipartimento di Matematica, Universit\`a
"La Sapienza" di Roma, Piazzale Aldo Moro 2, I-00185 Roma, Italy.
E-mail: dancona@mat.uniroma1.it}

\thanks{Luca Fanelli, Dipartimento di Matematica, Universit\`a
"La Sapienza" di Roma, Piazzale Aldo Moro 2, I-00185 Roma, Italy.
E-mail: fanelli@mat.uniroma1.it}

\begin{abstract}
    Given a one dimensional perturbed Schr\"odinger operator
    $H=-d^{2}/dx^{2}+V(x)$,
    we consider the associated wave operators
    $W_{\pm}$, defined as the strong $L^{2}$ limits
    $\lim_{s\to\pm\infty}e^{isH}e^{-isH_{0}}$.
    We prove that $W_{\pm}$ are bounded operators on $L^{p}$
    for all $1<p<\infty$, provided $(1+|x|)^{2}V(x)\in L^{1}$,
    or else $(1+|x|)V(x)\in L^{1}$ and 0 is not a resonance.
    For $p=\infty$ we obtain
    an estimate in terms of the Hilbert transform. Some applications
    to dispersive estimates for equations with variable rough
    coefficients are given.
\end{abstract}


\maketitle

\section{Introduction}\label{sec.introduc}

Let $H_{0}=-d^{2}/dx^{2}$ be the one-dimensional Laplace
operator on the line, and consider the perturbed operator
$H=H_{0}+V(x)$. For a potential $V(x)\in L^{1}(\R)$,
the operator $H$ can be realized uniquely as a selfadjoint operator
on $L^{2}(\R)$ with form domain $H^{1}(\R)$. The
absolutely continuous spectrum of
$H$ is $[0,+\infty[$, the singular spectrum is absent, and the
possible eigenvalues are all strictly negative and finite in number.
Moreover, the \emph{wave operators}
\begin{equation}\label{eq.wo}
    W_{\pm}f=L^{2}-\lim_{s\to\pm\infty}e^{isH}e^{-isH_{0}}f
\end{equation}
exist and are unitary from $L^{2}(\R)$ to the
absolutely continuous space $L^{2}_{ac}(\R)$ of $H$. A
very useful feature of $W_{\pm}$ is the
\emph{intertwining property}. If we denote
by $P_{ac}$ the projection of $L^{2}$ onto $L^{2}_{ac}(\R)$,
the property can be stated as follows:
for any Borel function $f$,
\begin{equation}\label{eq.twine}
    W_{\pm}f(H_{0})W_{\pm}^{*}=f(H)P_{ac}
\end{equation}
(see e.g. \cite{DeiTru}, \cite{CK}).

Thanks to \eqref{eq.twine}, one can reduce the study
of an operator $f(H)$, or more generally
$f(t,H)$, to the study of
$f(t,H_{0})$ which has a much simpler
structure. When applied to the operators
$e^{itH}$, $\frac{\sin(t\sqrt H)}{\sqrt H}$,
$\frac{\sin(t\sqrt{H+1})}{\sqrt{H+1}}$, this
method can be used to prove decay estimates for the Schr\"odinger,
wave and Klein-Gordon equations
\begin{equation*}
    iu_{t}-\Delta u+Vu=0,\qquad
    u_{tt}-\Delta u+Vu=0,\qquad
    u_{tt}-u_{xx}-\Delta u+u+Vu=0,
\end{equation*}
provided one has some control on the $L^{p}$
behaviour of $W_{\pm}$, $W^{*}_{\pm}$. Indeed, if
the wave operators are bounded on $L^{p}$,
the $L^{q}-L^{q'}$ estimates valid for the free
operators extend immediately to the perturbed
ones via the elementary argument
\begin{align*}
    \|e^{itH}P_{ac}f\|_{L^{q}}\equiv &
    \|W_{+}e^{itH_{0}}W_{+}^{*} f\|_{L^{q}}\\
    \leq & C\|e^{itH_{0}}W_{+}^{*} f\|_{L^{q}}
    \leq Ct^{-\alpha}\|W_{+}^{*} f\|_{L^{q'}}
    \leq Ct^{-\alpha}\| f\|_{L^{q'}}
\end{align*}

Such a program was developed systematically by K.Yajima
in a series of papers
\cite{Yajima95-waveopN},
\cite{Yajima95-waveopNeven},
\cite{Yajima99-waveop2D}
where he obtained
the $L^{p}$ boundedness for all $p$ of $W_{\pm}$,
under suitable assumptions on the potential $V$, for
space dimension $n\geq2$.
The analysis
was completed in the one dimensional case in Artbazar-Yajma
\cite{ArbYaj00} and Weder \cite{Weder99}.
We remark that in high dimension $n\geq4$ the
decay estimates obtained by this method are the best available
from the point of view of the assumptions on the potential;
only in low dimension $n\leq3$ more precise
results have been proved (see
\cite{GoldbergSchlag04},
\cite{Goldberg04}, \cite{RodnianskiSchlag04-disp},
\cite{Y05} and
\cite{DP}). We also mention \cite{GV05} for an interesting
class of related counterexamples.

In order to explain the results in more detail
we recall a few notions. The relevant potential
classes are the spaces
\begin{equation}\label{eq.L1ga}
    L^{1}_{\gamma}(\R)\equiv
     \{f\colon (1+|x|)^{\gamma}f\in L^{1}(\R)\}.
\end{equation}
Moreover, given a potential $V(x)$, the \emph{Jost functions} are
the solutions $f_{\pm}(\lambda,x)$ of the equation
$-f''+Vf=\lambda^{2}f$ satisfying the asymptotic conditions
$|f_{\pm}(\lambda,x)-e^{\pm\lambda x}|\to0$ as $x\to\pm\infty$. When
$V(x)\in L^{1}_{1}$, the solutions $f_{\pm}$ are uniquely defined
(\cite{DeiTru}). Now consider the Wronskian
\begin{equation*}
    W(\lambda)=
    f_{+}(\lambda,0)\partial_{x}f_{-}(\lambda,0)-
       \partial_{x}f_{+}(\lambda,0)f_{-}(\lambda,0).
\end{equation*}
The function $W(\lambda)$ is always different from zero for
$\lambda\neq0$, and it can only vanish at $\lambda=0$.
Then we say that 0 \emph{is a resonance for $H$} when
$W(0)=0$, and that it is not a resonance when $W(0)\neq0$.
The first one is also called the \emph{exceptional case}.

In \cite{Weder99} Weder proved that the wave operators
are bounded on $L^{p}$ for all $1<p<\infty$, provided
$V\in L^{1}_{\gamma}$ for $\gamma>5/2$.
The assumption can be relaxed to $\gamma>3/2$
provided 0 is not a resonance.
It is natural to conjecture that these conditions
may be sharpened, also in view of the $L^{\infty}-L^{1}$
decay estimate for the perturbed Schr\"odinger
equation proved by Goldberg and Schlag
\cite{GoldbergSchlag04} under the milder assumption
$\gamma=2$ in the general and $\gamma=1$ in
the nonresonant case.

Indeed, the main result of the present paper is the following:

\begin{thm}\label{thm.1}
    Assume $V\in L^1_1$ and 0 is not a resonance, or
    $V\in L^1_2$ in the general case. Then the wave operators
    $W_{\pm}, W^{*}_{\pm}$ can be extended to bounded
    operators on $L^{p}$ for all $1<p<\infty$. Moreover, in the
    endpoint $L^{\infty}$ case we have the estimate
    \begin{equation}\label{eq.stimaellep}
        \|W_\pm g\|_{L^\infty}\leq C\|g\|_{L^\infty}+
               C\|\HH g\|_{L^\infty},
    \end{equation}
    for all $g\in L^{\infty}\cap L^{p}$ for some $p<\infty$
    such that $\HH g\in L^{\infty}$,
    where $\HH$ is the Hilbert transform on $\R$;
    the conjugate operators $W^{*}_{\pm}$
    satisfy the same estimate.
\end{thm}

\begin{remark}\label{rem.hilb}
The appearence of the Hilbert trasform (see the beginning
of Section \ref{sec.bassenerg} for a quick reminder) at the
endpoint $p=\infty$ is not a surprise. Indeed, the very
precise analysis of Weder showed that the wave operator
can be decomposed as the sum of a multiple of the Hilbert
transform, plus a term bounded on $L^{\infty}$. Thus
a weaker estimate like \eqref{eq.stimaellep} is actually
optimal.

At the opposite endpoint $p=1$, we get an even weaker result
by duality (see Remark \ref{rem.int}). We conjecture that
by a suitable
modification of our methods a stronger bound
\begin{equation}\label{eq.stL1}
        \|W_\pm g\|_{L^1}\leq C\|g\|_{L^1}+
               C\|\HH g\|_{L^1}
\end{equation}
can be proved. Notice that \eqref{eq.stL1} is equivalent to
\begin{equation}\label{eq.stH1}
        \|W_\pm g\|_{L^1}\leq C\|g\|_{\HH_{1}}
\end{equation}
where $\HH_{1}$ is the Hardy space; by duality this would imply
\begin{equation}\label{eq.stBMO}
        \|W_\pm g\|_{BMO}\leq C\|g\|_{L^\infty}.
\end{equation}
A further evidence in this direction is that
the above estimates are a consequence of Weder's
decomposition,
of course under stronger assumptions on the potential.
\end{remark}

\begin{remark}\label{rem.GS}
Our proof is based on the improvement of some results
of Deift and Trubowitz \cite{DeiTru},
combined them with the
stationary approach of Yajima \cite{Yajima95-waveopN},
\cite{ArbYaj00}, and some precise Fourier analysis
arguments.
Quite inspirational has been the paper
\cite{GoldbergSchlag04}, both for showing there was room
for improvement in the assumptions on the potential, and
for the very effective harmonic analysis approach.
We mention in particular the idea of using Wiener's lemma
to estimate the $L^{1}$ norm of the Fourier transform of
a quotient, essential in Section \ref{sec.jost} below.
\end{remark}

\begin{remark}\label{rem.highfr}
In the proof of Theorem \ref{thm.1} we split as usual
the wave operator into high and low energy parts;
the high energy part is known to be easier to handle
since the resolvent is only singular at
frequency  $\lambda=0$. Here we can prove that
the high energy part is bounded on $L^{p}$ for all $p$,
including
the cases $p=1$ and $p=\infty$, under the weaker
assumption $V\in L^{1}(\R)$ (see
Section \ref{sec.altenerg} and Lemma \ref{lem.Wtilde}).
\end{remark}

\begin{remark}\label{rem.dt}
An essential step in the low energy estimate is a study of
the Fourier properties of the Jost functions; this kind of analysis
is classical (see \cite{AM}) and the fundamental estimates
were obtained by Deift and Trubowitz in \cite{DeiTru}.
In Section \ref{sec.jost} we improve their results
by showing that the $L^{1}$ norms of the Fourier transforms
of the Jost functions satisfy a linear bound as $|x|\to+\infty$
instead of an exponential one as in \cite{DeiTru}.
In the resonant case
we can prove a quadratic bound (see Lemmas \ref{lem.B},
\ref{lem.C} and Corollary \ref{cor.n}).
\end{remark}

\begin{remark}\label{rem.Wkp}
It is possible to continue the analysis and prove that the
wave operators are bounded on Sobolev spaces $W^{k,p}$,
under the additional assumption $V\in W^{k,1}$ (see
\cite{Weder99}), but we prefer not to pursue this question
here.
\end{remark}

Theorem \ref{thm.1} has several applications; here we
shall focus on the dispersive estimates for the
one dimensional Schr\"odinger
and Klein-Gordon equations with variable rough coefficients.

Consider first the initial value problem
\begin{equation}\label{eq.schrovariab}
    iu_t-a(x)u_{xx}+b(x)u_x+V(x)u=0,\qquad
    u(0,x)=f(x).
\end{equation}
Then we obtain the following decay result,
where the notation $f\in L^{2}_{1}$
means $(1+|x|)f\in L^{2}$. Notice that the case
$a=1$, $b=0$ is already considered in \cite{GoldbergSchlag04},
where actually the endpoint $L^{\infty}-L^{1}$ is reached.

\begin{prop}\label{prop.1}
    Assume $V\in L^{1}_{2}$,
    $a\in W^{2,1}(\R)$ and $b\in
    W^{1,1}(\R)$ with
    \begin{equation}\label{eq.i}
        a(x)\geq c_{0}>0\qquad
        a',b\in L^{2}_{1},\qquad
        a'',b'\in L^{1}_{2}
    \end{equation}
    for some constant $c_{0}$.
    Then the solution of the initial value problem
    \eqref{eq.schrovariab} satisfies
    \begin{equation}\label{eq.dispschro}
        \|P_{ac}u(t,\cdot)\|_{L^{q}}\leq
             Ct^{\frac1q-\frac12}
               \|f\|_{L^{q'}},\qquad
           2\leq q<\infty,\quad
           \frac1q+\frac1{q'}=1.
    \end{equation}
    The same result holds if $a=1$, $b=0$ and
    $V\in L^{1}_{1}$, provided 0 is not a resonance
    for $H$.
\end{prop}

\begin{remark}\label{rem.bad}
As discussed above, the case $q=\infty$ escapes this
method since the wave operator is not bounded on $L^{\infty}$.
It is however possible to recover the estimate also in this
case by the direct approach of \cite{GoldbergSchlag04}
\end{remark}

\begin{proof}
Define the functions
\begin{equation}\label{eq.csig}
    c(x)=\int_{0}^xa(s)^{-1/2}ds,\qquad
    \sigma(x)=a(x)^{1/4}
      \exp\left(\int_{0}^x\frac{b(s)}{2a(s)}\,ds\right),
\end{equation}
and apply the change of variables
\begin{equation}\label{eq.cambio}
    u(t,x)=\sigma(x)w(t,c(x)).
\end{equation}
Then the problem is transformed to
\begin{equation}\label{eq.cambioschro}
    iw_t(t,y)-w_{yy}+\widetilde V(y)w(t,y)=0,\qquad
    w(0,y)=\left.\frac f \sigma \right|_{c^{-1}(y)},
\end{equation}
where the potential $\widetilde V$ is defined by
\begin{equation}\label{eq.cambiopot}
    \widetilde V(c(x))=V(x)+
    \frac{1}{16a(x)}(2b(x)+a'(x))
    (2b(x)+3a'(x))-\frac14(2b(x)+a''(x)).
\end{equation}
It is elementary to check that $\widetilde V$ satisfies the
assumptions of Theorem \ref{thm.1}. Hence by the
intertwining property and the $L^{p}$ boundedness of
the wave operator for $H_{0}+\widetilde V$ we obtain
that $w(t,y)$ satisfies a dispersive estimate like
\eqref{eq.dispschro}. Coming back to the original variables
we conclude the proof.
\end{proof}

\begin{remark}\label{rem.burq}
The range of indices allowed in \eqref{eq.dispschro}
is sufficient to deduce the full set of Strichartz estimates,
as it is well known. It is interesting to compare this
with the result of Burq and Planchon \cite{BP} who proved
the Strichartz estimates for the variable coefficient
equation
\begin{equation*}
    iu_{t}-\partial_{x}(a(x)\partial_{x}u)=0
\end{equation*}
assuming only that $a(x)$ is of $BV$ class and
bounded from below.
\end{remark}

\begin{remark}\label{rem.bes}
In view of the next application,
we recall the definition of nonhomogeneous \emph{Besov
spaces}. Choose a Paley-Littlewood partition of unity, i.e.,
a sequence of smooth cutoffs $\phi_{j}\in C^{\infty}_{0}(\R)$
with $\sum_{j\geq0}\phi_{j}(\lambda)=1$ and
$\supp \phi_{j}=[2^{j-1},2^{j+1}]$,
$\supp \phi_{0}=[-2,2]$. Then the $B^{s}_{p,r}$ Besov norm
is defined by
\begin{equation*}
    \|g\|_{B^{s}_{p,r}}^{r}\equiv
        \sum_{j\geq0} 2^{jsr}\|\phi_{j}
               (\sqrt H_{0})g\|_{L^{p}}^{r}
\end{equation*}
with obvious modification for $r=\infty$. It is then natural to
define the \emph{perturbed Besov norm} corresponding to the
selfadjoint operator $H=H_{0}+V$ as
\begin{equation*}
    \|g\|_{B^{s}_{p,r}(V)}^{r}\equiv
        \sum_{j\geq0} 2^{jsr}\|\phi_{j}
               (\sqrt H)g\|_{L^{p}}^{r}.
\end{equation*}
Now, from the $L^{p}$ boundedness of the wave operators
and the intertwining property in the form
\begin{equation*}
    \phi_{j}(\sqrt H)W_{\pm}
       =W_{\pm}\phi_{j}(\sqrt {H_{0}})
\end{equation*}
we obtain immediately the Besov space bounds
\begin{equation}\label{eq.besoW}
     \|W_{\pm}f\|_{B^{s}_{p,r}(V)}\leq
     C\|f\|_{B^{s}_{p,r}},\qquad
    \|W^{*}_{\pm}f\|_{B^{s}_{p,r}}\leq
     C\|f\|_{B^{s}_{p,r}(V)}.
\end{equation}
\end{remark}

We now consider the initial value problem
for the one dimensional Klein-Gordon equation
\begin{equation}\label{eq.KGvariab}
    u_{tt}-a(x)u_{xx}+u+b(x)u_x+V(x)u=0,\qquad
    u(0,x)=0,\quad u_t(0,x)=g(x).
\end{equation}
Our second application is the following:

\begin{prop}\label{prop.2}
    Assume $V,a,b$ are as in Proposition \ref{prop.1}.
    Then the solution of the initial value problem
    \eqref{eq.KGvariab} satisfies
    \begin{equation}\label{eq.dispKG}
        \|P_{ac}u(t,\cdot)\|_{L^{q}}\leq
             Ct^{\frac1q-\frac12}
               \|f\|_{B^{\frac 12-\frac{3}q}_{q',q}(V)},\qquad
           2\leq q<\infty,\quad
           \frac1q+\frac1{q'}=1.
    \end{equation}
    The same result holds if $a=1$, $b=0$ and
    $V\in L^{1}_{1}$, provided 0 is not a resonance
    for $H$.
\end{prop}

\begin{proof}
For the free Klein-Gordon equation
\begin{equation*}
    u_{tt}-u_{xx}+u=0,\qquad
    u(0,x)=0,\quad u_t(0,x)=g(x).
\end{equation*}
we know that the solution
satisfies the energy estimate and the dispersive estimate,
which we can write in the form
\begin{equation*}
    \|u(t,\cdot)\|_{L^{2}}\leq C\|g\|_{H^{-1}},\qquad
    \|u(t,\cdot)\|_{L^{\infty}}\leq C\,t^{-\frac12}
    \|g\|_{B^{1/2}_{1,1}}.
\end{equation*}
By real interpolation this implies
\begin{equation*}
       \|u(t,\cdot)\|_{L^{q}}\leq
             Ct^{\frac1q-\frac12}
               \|f\|_{B^{\frac 12-\frac{3}q}_{q',q}},\qquad
           2\leq q\leq\infty,\quad
           \frac1q+\frac1{q'}=1.
\end{equation*}
Consider now the variable coefficient problem
\eqref{eq.KGvariab}; applying the same change of
variables \eqref{eq.csig}-\eqref{eq.cambio} as
above, we transform the problem into
\begin{equation}\label{eq.cKG}
    w_{tt}(t,y)-w_{yy}+w+\widetilde V(y)w(t,y)=0,\qquad
    w(0,y)=0,\quad
    w_{t}(0,y)=\left.\frac g \sigma \right|_{c^{-1}(y)}
\end{equation}
with $\widetilde V$ as in \eqref{eq.cambiopot}. Applying
as above the intertwining property, and using the bounds
\eqref{eq.besoW}, we conclude the proof.
\end{proof}

The rest of the paper is devoted to the proof of Theorem
\ref{thm.1}. We first analyze the
high energy part, in Section \ref{sec.altenerg};
Section \ref{sec.jost} contains a detailed study
of the Fourier properties of the Jost functions, necessary
for the analysis of the low energy part which is the
subject of Section \ref{sec.bassenerg}.

\begin{ack}
We would like to thank Kenji Yajima for several very useful
discussions concerning the subject of this paper.
\end{ack}

\section{The high energy analysis}\label{sec.altenerg}

In the estimate of the high frequency part of the
wave operator we shall use the standard representation
as a distorted Fourier transform; considering e.g. the operator
$W_{-}$, we have
\begin{equation}\label{eq.W-}
    W_-g(x)=\frac{1}{2\pi}\int_{-\infty}
    ^{+\infty}\left(\int_{-\infty}^{+\infty}
    \varphi(\lambda,x)e^{-i\lambda y}
    \,d\lambda\right)g(y)\,dy,
\end{equation}
where the generalized eigenfunction $\varphi(\lambda,x)$
is defined as the solution to the Lippman-Schwinger equation
\begin{equation}\label{eq.lip-sch}
    \varphi(\lambda,x)=e^{i\lambda x}+
    \frac{1}{2i|\lambda|}\int e^{i|\lambda|
    \cdot|x-y|}V(y)\varphi(\lambda,y)\,dy,
    \qquad \lambda,x\in\R
\end{equation}
(see e.g. \cite{ArbYaj00}, \cite{Weder99}). An
equivalent form of the equation \eqref{eq.lip-sch} is
the following:
\begin{equation}\label{eq.lip-sch2}
    \varphi(\lambda,x)=e^{i\lambda x}-
    R_0(\lambda^2+i0)V
    \varphi(\lambda,x).
\end{equation}

We recall that the free resolvent $R_0(z)=(-\Delta-z)^{-1}$
admits the explicit representation
\begin{equation}\label{eq.R0zeta}
    R_0(z^2)f(x)=\frac{1}{2i}\int \frac{
    e^{iz|x-y|}}{z}f(y)\,dy,
\end{equation}
for $z\not\in[0,+\infty[$. Moreover, the limits
\begin{equation}\label{eq.lap}
    R_0(\lambda\pm i0)=\lim_{\epsilon\to0}
    R_0(\lambda\pm i\epsilon)
\end{equation}
exist in the norm of bounded operators from the
weighted $L^2_{1/2+\epsilon}$ to the
weighted  $L^2_{-1/2-\epsilon}$
spaces, for any $\lambda\in ]0,\infty[$
(see e.g. \cite{agmon}).
Thus we have the explicit formula
\begin{equation}\label{eq.R0lambda}
    R_0(\lambda\pm i0)f(x)=\frac{1}{2i}
    \int\frac{e^{\pm i\lambda|x-y|}}{\lambda}
    f(y)\,dy,
\end{equation}
for any $\lambda>0$ and $f$ at least in $L^1$. The
strong singularity at $\lambda=0$ is the main source of
difficulties in the study of the wave operator.

The perturbed resolvent
$R_V(z)=(-\Delta+V-z)^{-1}$ is related to $R_0$ by the identity
\begin{equation}\label{eq.RV}
    R_V=R_0(I+VR_0)^{-1}.
\end{equation}
We recall that under the assumption $V\in L^1_1$
the limiting absorption principle \eqref{eq.lap} holds also
for $R_V$ (see \cite{BRV}, \cite{PdaLf05}).

By the representation \eqref{eq.R0lambda} it is clear that
for $\lambda\geq\lambda_{0}=\|V\|_{L^1}$
the operator
$R_{0}V$ is bounded on $L^{\infty}$ with norm
\begin{equation*}
    \|R_0(\lambda^2+i0)V\|_{
    \mathcal L(L^\infty)}\leq\frac12.
\end{equation*}
In particular, for $\lambda$ large enough,
$I+R_0(\lambda ^2+i0)V$ can be inverted by
a Neumann series, the solution $\phi(\lambda,x)$
of \eqref{eq.lip-sch2}
is well defined and it can be represented
by a uniformly convergent series
\begin{equation}\label{eq.serie}
    \varphi(\lambda,x)=\sum_{n\geq0}(-1)^{n}
    \left(R_0(\lambda^2+i0)V\right)^n
    e^{ikx},\qquad
    |\lambda|\geq\lambda_0:={\|V\|_{L^1}},\quad
    x\in\R.
\end{equation}
Now take a smooth cutoff
function $\Phi\in C^\infty(\R^{+})$ such that
\begin{equation*}
    0\leq \Phi\leq1,\qquad
   \Phi(\lambda^{2})=0\quad\text{for}\quad
    0\leq\lambda^{2}\leq\lambda_{0},\qquad
   \Phi(\lambda^{2})=1\quad\text{for}\quad
    \lambda^{2}\geq\lambda_{0}+1
\end{equation*}
and consider the high energy part of the wave operator
\begin{equation*}
    W_{-} \Phi(H_{0})g(x)=
    \frac{1}{2\pi}\int_{-\infty}
    ^{+\infty}\int_{-\infty}^{+\infty}
    \varphi(\lambda,x)e^{-i\lambda y}
    g(y) \Phi(\lambda^{2})\,d\lambda dy.
\end{equation*}
We split this operator into positive and negative  frequencies,
i.e., writing
\begin{equation*}\label{eq.posneg}
    \chi(\lambda)=
    \begin{cases}
        \Phi(\lambda^{2})&\text{for $\lambda>0$,}\\
        0&\text{for $\lambda\leq0$}
    \end{cases}
    \qquad
    \psi(\lambda)=
    \begin{cases}
        \Phi(\lambda^{2})&\text{for $\lambda<0$,}\\
        0&\text{for $\lambda\geq0$}
    \end{cases}
\end{equation*}
we define the operators
\begin{equation}\label{eq.AA}
    Ag(x)=
    \frac{1}{2\pi}\int_{-\infty}
    ^{+\infty}\int_{-\infty}^{+\infty}
    \varphi(\lambda,x)e^{-i\lambda y}
    g(y) \chi(\lambda)\,d\lambda dy
\end{equation}
and
\begin{equation}\label{eq.BB}
    Bg(x)=
    \frac{1}{2\pi}\int_{-\infty}
    ^{+\infty}\int_{-\infty}^{+\infty}
    \varphi(\lambda,x)e^{-i\lambda y}
    g(y) \psi(\lambda)\,d\lambda dy.
\end{equation}
In the following we shall study the positive part
$Ag$; clearly the estimate of the negative piece $Bg$ is completely
analogous.
By \eqref{eq.W-} and \eqref{eq.serie}, the integral kernel
$K(x,y)$ of $A$
\begin{equation}\label{eq.W2}
    Ag(x)=\frac{1}{2\pi}\int
    K(x,y)g(y)\,dy,
\end{equation}
can be represented as
\begin{equation}\label{eq.K}
    K(x,y)=\sum_{n\geq0}(-1)^{n}
    \int\left(
    R_0(\lambda^2+i0)V\right)^ne^{i\lambda
    (x-y)}\chi(\lambda)\,d\lambda.
\end{equation}
We shall estimate the terms of the series \eqref{eq.K}
separately. Notice that for $n\geq2$ we can write
\begin{equation}\label{eq.Kn}
    K_n(x,y)=\left(\frac i2\right)^n\int
    \!\!\dots\!\!
    \int\frac{\chi(\lambda)}{\lambda^n}
    e^{i\lambda(|x-y_1|+|y_1-y_2|+\dots+|y_
    {n-1}-y_n|+y_n-y)}
    \prod_{j=1}^nV(y_j)
    \,dy_1\dots\,
    dy_n\,d\lambda.
\end{equation}
On the other hand, for $n=0,1$ we have the formal expressions
\begin{equation}\label{eq.K0}
    K_0(x,y)=\int e^{i\lambda(x-y)}\chi
    (\lambda)\,d\lambda,
\end{equation}
\begin{equation}\label{eq.K1}
    K_1(x,y)=\frac i2\int\!
    \int\frac{\chi
    (\lambda)}{\lambda}e^{i\lambda
    (|x-y_1|+y_1-y)}V(y_1)\,dy_1\,d\lambda
\end{equation}
which can defined precisely by adding a cutoff on
$[0,L]$ and then sending $L\to+\infty$
(see below).
Denoting by $A_{n}$ the operator
with kernel $K_{n}(x,y)$, we have
\begin{eqnarray}
    Ag(x) & = & \frac{1}{2\pi}
    \left[\int
    K_0(x,y)g(y)\,dy+\int K_1(x,y)g(y)\,dy
    -\sum_{n\geq2}(-1)^{n}\int K_n(x,y)
    g(y)\,dy\right]\nonumber
    \\ \  & = & \frac{1}{2\pi}
    \left(
    A_0g(x) - A_1g(x)
    +\sum_{n\geq2}(-1)^{n}
    A_ng(x)\right).
    \label{eq.Wtilde}
\end{eqnarray}

Then we have:

\begin{lem}\label{lem.Wtilde}
    Assume $V\in L^1(\R)$ and
    let $0\leq\Phi\leq1$ be a smooth function
    such that $\Phi(\lambda^{2})=0$
    for $\lambda^{2}<\|V\|_{L^{1}}$ and
    $\Phi(\lambda^{2})=1$
    for $\lambda^{2}>\|V\|_{L^{1}}+1$.
    Then the high energy parts of the wave
    operators $W_{\pm}$ are bounded on $L^{p}$
    for all $1\leq p\leq\infty$:
    \begin{equation}\label{eq.stimatilde}
        \|W_{\pm}\Phi(H_{0})g\|_{L^p}
        \leq C\|g\|_{L^p}.
    \end{equation}
    The same holds for the conjugate operators
    $\Phi(H_{0})W^{*}_{\pm}$.
\end{lem}

\begin{proof}
By standard duality arguments, it will be sufficient to prove
the estimates for $p=\infty$; since the proof is completely
analogous for
any of the four operators $W_{\pm}$, $W_{\pm}^{*}$,
we shall consider only $W_{-}$. By the discussion above,
we see that it is sufficient to estimate the operator $A$
defined in \eqref{eq.W2}-\eqref{eq.Wtilde}.

We shall estimate each term $A_{n}$ in the series
\eqref{eq.Wtilde} separately.
For the term $A_0$, we can write by
\eqref{eq.K0}
\begin{eqnarray*}
    A_0g(x) & = & \int\left(\int
    e^{i\lambda(x-y)}\chi(\lambda)g(y)\,d\lambda
    \right)\,dy \\ \  & = &
    \int\!\int
    e^{i\lambda(x-y)}[1-(1-\chi(\lambda))]
    g(y)\,d\lambda\,dy \\ \  & = &
    g(x)-[\widehat{(1-\chi)}*g](x),
\end{eqnarray*}
(recall the notations $\hat h=\mathcal{F}h$
for the Fourier transform of a function $h$)
whence we obtain
\begin{equation}\label{eq.stima0}
    \|A_0g\|_{L^\infty}\leq
    \left(1+\|\widehat{(1-\chi)}\|_{L^1}
    \right)\|g\|_{L^\infty}\leq C_0
    \|g\|_{L^\infty}.
\end{equation}
Consider now the term $A_1$, which by \eqref{eq.K1} can
be written formally
\begin{equation}\label{eq.W1}
    A_1g(x)=\frac i2\int
    \left(\int\left(
    \int\frac{\chi(\lambda)}\lambda
    e^{i\lambda(|x-z|+z-y)}V(z)g(y)\,dz
    \right)\,d\lambda\right)\,dy.
\end{equation}
More precisely, fixed a function
$\psi(\lambda)\in C^{\infty}_{c}$
equal to 1 on $[-1,1]$ and vanishing outside $[-2,2]$,
we define the truncated operators
\begin{equation}\label{eq.W1L}
    A_{1,L}g=\frac i2
    \int
    \left(\int\left(
    \gamma_{L}(\lambda)
    e^{i\lambda(|x-z|+z-y)}V(z)g(y)\,dz
    \right)\,d\lambda\right)\,dy,
\end{equation}
where
\begin{equation*}
    \gamma_{L}(\lambda)=\frac1\lambda\chi(\lambda)
            \psi_{L}(\lambda),\qquad
    \psi_{L}(\lambda)\equiv\psi(\lambda/L)
\end{equation*}
We claim that the operators $A_{1,L}$
are uniformly bounded on $L^{\infty}$, and that
for each $g\in L^{\infty}$ the limit
\begin{equation}\label{eq.WLtoW1}
    A_1g= \lim_{L\to+\infty}A_{1,L}
\end{equation}
exist in the norm of $L^{\infty}$. To prove this, we notice that by
Fubini's theorem  \eqref{eq.W1L} can be rewritten as
\begin{equation}\label{eq.W1L2}
    A_{1,L}g=\frac i2
    \int\int\hat
    \gamma_{L}(|x-z|+z-y)
    V(z)g(y)\,dz\,dy,
\end{equation}
It is clear that the claim follows as soon as we can prove
that $\hat\gamma_{L}$ converges in
$L^{1}(\R)$ when $L\to+\infty$:
indeed, we have
\begin{equation*}
    \|A_{1,L}g-A_{1,M}g\|_{L^{\infty}}\leq
         \|V\|_{L^{1}}
         \|\hat\gamma_{L}-\hat\gamma_{M}\|_{L^{1}}
         \|g\|_{L^{\infty}}.
\end{equation*}
To prove the claim, decompose $\gamma_{L}$ as follows:
\begin{equation}\label{eq.gaL}
    \gamma_{L}(\lambda)=
      \psi_{L}(\lambda)\cdot
      \left(
      \frac{\lambda}{1+\lambda^2}+
      (\chi-1)\frac{\lambda}{1+\lambda^2}+
      \frac{\chi(\lambda)}{\lambda(1+\lambda^2)}
      \right).
\end{equation}
The function
\begin{equation}\label{eq.partL}
    \eta(\lambda)=
      \frac{\lambda}{1+\lambda^2}+
      (\chi-1)\frac{\lambda}{1+\lambda^2}+
      \frac{\chi(\lambda)}{\lambda(1+\lambda^2)}
\end{equation}
has a Fourier transform in $L^{1}$; this follows immediately
from the standard formula
\begin{equation}\label{eq.psihat}
    \widehat{\frac{\lambda}{1+\lambda^2}}=
      C\frac{\xi}{|\xi|}e^{-|\xi|}
\end{equation}
and the fact that the last term in \eqref{eq.partL} is
smooth and decays faster than $|\lambda|^{-3}$.
Since $\hat\psi_{L}$ is a $\delta$-sequence,
we conclude that
$\hat\gamma_{L}=\hat\psi_{L}*\hat\eta$ converges
to $\hat\eta$ in $L^{1}(\R)$. As a consequence,
$A_{1,L}g$ converge uniformly to
\begin{equation*}
    A_{1}g\equiv\frac i2\int\int\hat\eta(|x-z|+z-y)V(z)g(y)dzdy
\end{equation*}
which is then a bounded operator on $L^{\infty}$:
\begin{equation}\label{eq.stima1}
    \|A_1g\|_{L^\infty}\leq
    \|V\|_{L^1}\|\hat\eta\|_{L^1}\|g\|
    _{L^\infty}.
\end{equation}

To conclude the proof,
it remains to estimate the operators $A_n$ for $n\geq2$. By the
explicit formula \eqref{eq.Kn} we obtain
\begin{eqnarray*}
    A_ng(x) & = & \left(\frac i2\right)^n
    \!\int\!\!\dots
    \!\!\int\psi_n(\lambda)
    e^{i\lambda(|x-y_1|+|y_1-y_2|+\dots+|y_
    {n-1}-y_n|+y_n-y)}\times
    \\ \  & \  & \times
    \prod_{j=1}^nV(y_j)g(y)
    \,dy_1\dots\,
    dy_n\,d\lambda\,dy,
\end{eqnarray*}
where $\psi_n(\lambda):=\chi(\lambda)/\lambda^n$. By Fubini's
Theorem this can be written
\begin{eqnarray*}
    A_ng(x) & = & \left(\frac
    i2\right)^n\!
    \int\!\!\dots
    \!\!\int\hat\psi_n(|x-y_1|+|y_1-y_2|
    +\dots+|y_{n-1}-y_n|+y_n-y)\times
    \\ \  & \  & \times
    \!\prod_{j=1}^nV(y_j)g(y)
    \,dy_1\dots
    dy_n\,dy,
\end{eqnarray*}
and then we immediately get the inequality
\begin{equation}\label{eq.quasistiman}
    \| A_ng(x)\|_{L^\infty}
    \leq \frac 1{2^n}\|V\|_{L^1}^n
   \|\hat\psi_n\|_{L^1}
    \|g\|_{L^\infty}.
\end{equation}
To compute the norm of $\hat\psi_{n}$,
introduce the scaling operators $S_{h}$
defined as $S_{h}g(x)=g(hx)$; then
writing
\begin{equation*}
    \chi_0(\lambda):=\chi(\lambda\cdot\lambda_{0}),
    \qquad \lambda_{0}=\|V\|_{L^{1}},
\end{equation*}
we have
\begin{equation*}
    \psi_n(\lambda)=\lambda_0^{-n}\cdot S_{1/\lambda_0}
    \left(\frac{\chi_0(\lambda)}{
    \lambda^n}\right)
\end{equation*}
and hence
\begin{eqnarray*}
    \|\hat\psi_n\|
    _{L^1} & = & \lambda_0^{-n}
    \|\mathcal FS_{\frac
    1{\lambda_0}}
    \left({\chi_0}/{\lambda^n}
    \right)\|_{L^1}=
    \lambda_0^{-n}
    \|\lambda_0S_{\lambda_0}\mathcal F
    \left({\chi_0}/{\lambda^n}
    \right)\|_{L^1}
    \\ \  & = &
    \lambda_0^{1-n}
    \|\mathcal F
    \left({\chi_0}/{\lambda^{n}}
    \right)\|_{L^1}
    \leq
    C\lambda_0^{1-n}
    \|\langle\xi\rangle^{2}\mathcal F
    \left({\chi_0}/{\lambda^{n}}
    \right)\|_{L^\infty}
    \\ \  & \leq &
    C\lambda_0^{1-n}
    \|(1-\Delta)
    \left({\chi_0}/{\lambda^{n}}
    \right)\|_{L^1}
    \leq C_0 n^{2}\lambda_0^{1-n}
    \equiv C_0n^{2}\|V\|_{L^{1}}^{1-n}
\end{eqnarray*}
for some constant $C_{0}$ independent of $n$ and $\lambda_{0}$
This inequality together with \eqref{eq.quasistiman} gives
\begin{equation}\label{eq.stiman}
    \|A_ng(x)\|_{L^\infty}
    \leq C_{0}\frac{n^{2}}{2^n}\|V\|_{L^{1}}
    \|g\|_{L^\infty}.
\end{equation}

By the estimates \eqref{eq.stima0}, \eqref{eq.stima1},
\eqref{eq.stiman} and by formula \eqref{eq.Wtilde}
we conclude the proof of the Lemma.
\end{proof}

\section{Fourier properties of the Jost Functions}\label{sec.jost}

Throughout this section we shall assume
that $V\in L^{1}_{1}(\R)$ (at least).

The \emph{Jost functions} $f_\pm(z,x)$
are defined as the solutions of
\begin{equation}\label{eq.jost}
    -f_\pm^{''}(z,x)+V(x)f_\pm(z,x)
    =z^2f_\pm(z,x)
\end{equation}
satisfying the asymptotic conditions
\begin{equation}\label{eq.jostcond}
    \left|f_\pm(z,x)-e^{\pm izx}\right|
    \to0,
\end{equation}
for $x\to\pm\infty$. It is well known (see \cite{DeiTru}) that
$f_\pm(\lambda,x)$ are well defined for all $\lambda,x\in\R$.
Using the Jost functions it is possible to represent the kernel
of the perturbed resolvent $R_{V}(\lambda^{2}\pm i0)$;
indeed, writing
\begin{equation*}
    R_{V}(\lambda^{2}\pm i0)g=\int K_{\pm}(x,y)g(y)dy,
\end{equation*}
one has
\begin{equation}\label{eq.jostRV}
   K_{\pm}(x,y)=\frac{1}{2\pi i}
   \begin{cases}
      \displaystyle
      \frac{f_+(\pm\lambda,y)f_-(\pm\lambda,x)}
                 {W(\pm\lambda)} & \text{for}\ x<y,
            \\ \\
      \displaystyle
      \frac{f_+(\pm\lambda,x)f_-(\pm\lambda,y)}
                {W(\pm\lambda)} & \text{for}\ x>y;
   \end{cases}
\end{equation}
here
$$W(\lambda)=
f_+(\lambda,0)\cdot \partial_{x}f_- (\lambda,0) -
\partial_{x}f_-(\lambda,0)\cdot f_+(\lambda,0)$$
denotes the
Wronskian of $f_+$ and $f_-$. It is always true
(see \cite{DeiTru}) that
$W(\lambda)\neq0$ for any $\lambda\neq0$;
thus the only possible zero of the Wronskian
is at $\lambda=0$, and when $W(0)=0$ we say that
0 is a {\it resonance} for $-\Delta+V$.

The modified Jost functions $m_\pm$ are defined via
\begin{equation}\label{eq.m}
    f_\pm(\lambda,x)=e^{\pm i\lambda x}
    m_\pm(\lambda,x);
\end{equation}
equivalently, the functions $m_\pm$ can be
characterized as
the unique solutions of the equations
\begin{equation}\label{eq.msolve}
    m_\pm''(\lambda,x)\pm2i\lambda
    m_\pm'(\lambda,x)=V(x)m_\pm(\lambda,x)
\end{equation}
satisfying the asymptotic conditions
\begin{equation}\label{eq.mcond}
    m_\pm(\lambda,x)\to1\quad\text{for}\quad
    x\to\pm\infty.
\end{equation}
Moreover, we can also obtain $m_\pm(\lambda,x)$ as
the unique solutions of the Volterra integral equations
\begin{equation}\label{eq.volterra}
    m_\pm(\lambda,x)=1\pm\int_x^{+\infty}
    D_\lambda(\pm(t-x))V(t)m_\pm(
    \lambda,t)\,dt,
\end{equation}
where
\begin{equation}\label{eq.Dlambda}
    D_\lambda(x):=\int_0^xe^{2i\lambda t}
    \,dt=\frac{e^{2i\lambda x}-1}{2i\lambda}.
\end{equation}
The functions $m_{\pm}(\lambda,x)$ have a rich
set of properties, studied in detail in \cite{DeiTru}.
Here we shall only need the following basic facts:
\begin{itemize}
\item
    if $V\in L^1_1$, then $m_\pm(\lambda,x)\in
    C(\R^2)$;
\item
    if $V\in L^1_2$, then
    $m_\pm(\lambda,x)\in C^1(\R^2)$ and
    $\frac\lambda{W(\lambda)}\in C(\R)$.
\end{itemize}


In scattering theory an essential role is played by
the Fourier transform w.r. to $\lambda$ of the
functions $m_\pm-1$, which are usually written in the form
\begin{equation}\label{eq.B}
    B_\pm(\xi,x)=
    \int_\R e^{-2i\lambda\xi}
    \left(
    m_\pm(\lambda,x)-1\right)d\lambda.
\end{equation}
(notice the factor 2 in the exponential).
For each $x\in\R$ the function $B_{+}(\xi,x)$ is well
defined, real valued, belongs
to $L^{2}(\R)$ and actually vanishes for $\xi<0$; this
means that $m_{+}(\cdot,x)-1$ belongs to the
\emph{Hardy space} $H^{2+}$ (see \cite{DeiTru}
for details). Analogously, $B_{-}(\xi,x)$ belongs
to $L^{2}(\R)$ and vanishes for $\xi>0$, i.e.,
$m_{-}(\cdot,x)-1\in H^{2-}$.

If we take the Fourier transform of equation \eqref{eq.volterra},
we obtain that $B_{+}(\xi,x)$ satisfies the
\emph{Marchenko equation}
\begin{equation}\label{eq.marchenko}
    B_+(\xi,x)=\int_{x+\xi}
    ^{\infty}V(t)\,dt
    +\int_0^\xi\,dz\int_{x+\xi-z}^{\infty}
    V(t)B_+(z,t)\,dt
\end{equation}
while $B_{-}(\xi,x)$ satisfies the symmetric equation
\begin{equation}\label{eq.marchenkom}
    B_-(\xi,x)=\int_{-\infty}^{x+\xi}
    V(t)\,dt
    +\int_\xi^{0}\,dz\int_{-\infty}^{x+\xi-z}
    V(t)B_-(z,t)\,dt
\end{equation}

The functions $B_\pm(\xi,x)$ have many additional properties
of boundedness and regularity; however we shall only be
concerned here with the properties of the $L^{1}$ norms
$\|B_{\pm}(\cdot,x)\|_{L^{1}}$.
Writing for $x\in\R$
\begin{equation*}
    \eta(x)=\int_{x}^{\infty}|V(t)|dt,\qquad
    \gamma(x)=\int_x^\infty(t-x)|V(t)|dt
       \equiv\int_{x}^{\infty}\int_{y}^{\infty}|V(t)|dtdy,
\end{equation*}
the well-known estimate of Deift and Trubowitz is
the following:

\begin{lem}\label{lem.lemma3DT}
    Assume $V\in L^{1}_{1}$. Then, for all $\xi,x\in\R$,
    the solution $B_+(\xi,x)$ to \eqref{eq.marchenko}
    is well defined and satisfies the estimates
    \begin{equation}\label{eq.DTbase}
        |B_{+}(\xi,x)|\leq e^{\gamma(x)}\eta(\xi+x),\qquad
        |\partial_{x}B_{+}(\xi,x)+V(x+\xi)|\leq
            e^{\gamma(x)}\eta(x+\xi).
    \end{equation}
    In particular, $B(\cdot,x)$ is in $L^1\cap L^\infty$
    for any $x$ and
    \begin{equation}\label{eq.stimeBDT}
        \|B_+(\cdot,x)\|_{L^1}\leq
            e^{\gamma(x)}\gamma(x),\qquad
        \|\partial_{x} B_+(\cdot,x)\|_{L^1}\leq
            \eta(x)+e^{\gamma(x)}\gamma(x).
    \end{equation}
\end{lem}

The function $B_{-}$ has similar properties, with the behaviours
at $\pm\infty$ reversed.
Notice that $\gamma(x)\leq\|V\|_{L^1_1}$ for $x\geq0$,
while for negative $x$ the behaviour of $\gamma(x)$ is
\begin{equation*}
    \gamma(x)\sim |x|\cdot\|V\|_{L^{1}}+
         \|V\|_{L^{1}_{1}},\quad x\to-\infty.
\end{equation*}
In other words, the estimate shows that
$\|B_+(\cdot,x)\|_{L^1}$ is bounded
by a constant depending on $\|V\|_{L^{1}_{1}}$ for $x>0$,
but it gives only an exponential bound
for negative $x$. A similar estimate holds for the function
$B_{-}$, exchanging the behaviours as $x\to+\infty$ and $x\to-\infty$.

A crucial tool in the study of the low energy case will be
an essential improvement of the \eqref{eq.stimeBDT}: indeed,
we can prove that the norm of $B_{+}$
(resp. $B_{-}$) has at most a linear growth as
$x\to-\infty$ (resp. $x\to+\infty$).

\begin{lem}\label{lem.B}
    Assume $V\in L^1_1$; then the functions
    $B_\pm(\xi,x)$ satisfy the estimates
    \begin{equation}\label{eq.stimaB}
        \|B_\pm(\cdot,x)\|_{L^1}
        \leq C
        \quad\text{for $\pm{x}\geq0$},
        \qquad
        \|B_\pm(\cdot,x)\|_{L^1}
        \leq C\langle x
        \rangle
        \quad\text{for $\pm{x}\leq0$}
    \end{equation}
    for some constant $C$ depending on $\|V\|_{L^{1}_{1}}$.
\end{lem}

\begin{proof}
We prove the result for $B_+$, the proof for $B_-$ is
identical. The behaviour for positive $x$ is already
contained in the Deift-Trubowitz estimate.
Now, starting from the Marchenko equation
\eqref{eq.marchenko},
we integrate with respect
to $\xi$ from 0 to $\infty$ (recall that $B_{\pm}$ vanish
for $\xi<0$) and we have
\begin{eqnarray*}
    \|B_+(\cdot,x)\|_{L^1} & \leq &
    \|V\|_{L^1_{1}}+
    |x|\cdot\|V\|_{L^1}+\int_0^\infty\,d\xi\int
    _0^\xi\,dz\int_{x+\xi-z}^{\infty}
    |V(t)|\cdot|B_+(z,t)|\,dt \\ \  & \leq &
    \xx\cdot\|V\|_{L^1_{1}}+\int_0^\infty \,d\xi\int
    _{-\infty}^\xi\,dz\int_{x+\xi-z}^{\infty}
    |V(t)|\cdot|B_+(z,t)|\,dt.
\end{eqnarray*}
We set $z':=\xi-z$ and exchange the order of integration: in this
way we obtain
\begin{eqnarray}
    \|B_+(\cdot,x)\|_{L^1} & \leq &
    \xx\cdot\|V\|_{L^1_{1}}+\int_x^\infty\,dt\int_0^{t-x}
    |V(t)|\cdot\|B_+(\cdot,t)\|
    _{L^1}\,dz'\nonumber \\ \  & = &
    \xx\cdot\|V\|_{L^1_{1}}+\int_x^\infty|V(t)|\cdot(t-x)\cdot
    \|B_+(\cdot,t)\|_{L^1}\,dt.
    \label{eq.quasigronw}
\end{eqnarray}
Now we remark that
\begin{equation*}
    \int_x^\infty t|V(t)|\cdot
    \|B_+(\cdot,t)\|_{L^1}dt
    \leq \int_0^\infty t|V(t)|\cdot
    \|B_+(\cdot,t)\|_{L^1}dt
\end{equation*}
which is obvious when $x>0$ and is also evident for $x<0$
since the integral from $x$ to 0 is negative. Using the
Deift-Trubowitz estimate \eqref{eq.stimeBDT} we see that
$$\|B_{+}(\cdot,t)\|\leq C_{0}=C_{0}(\|V\|_{L^{1}_{1}})
\quad\text{for $t>0$,}$$
and hence in conclusion
\begin{equation*}
    \int_x^\infty t|V(t)|\cdot
    \|B_+(\cdot,t)\|_{L^1}dt
    \leq  C_{1} \equiv C_{1}(\|V\|_{L^{1}_{1}})
    \quad\text{for all $x\in\R$.}
\end{equation*}
Thus inequality \eqref{eq.quasigronw} gives
\begin{equation*}
     \|B_+(\cdot,x)\|_{L^1} \leq
     \xx\cdot\|V\|_{L^1_{1}}+C_{1}(\|V\|_{L^{1}_{1}})+|x|
     \int_x^\infty|V(t)|\cdot\|B_{+}(\cdot,t)\|_{L^{1}}dt
\end{equation*}
which implies
\begin{equation}\label{eq.gronwall}
    \frac{1}{\langle x\rangle}
    \|B_+(\cdot,x)\|_{L^1}
    \leq C_2(\|V\|_{L^{1}_{1}})+\int_x^\infty\langle t\rangle
    |V(t)|\frac{\|B_+(\cdot,t)\|_{L^1}}
    {\langle t\rangle}\,dt.
\end{equation}
Applying Gronwall's lemma
for $x<0$ we finally obtain the required estimate
\begin{equation*}
    \|B_+(\cdot,x)\|_{L^1}\leq
    C_3(\|V\|_{L^{1}_{1}})\cdot\langle x\rangle.
\end{equation*}
\end{proof}

In the resonant case $W(0)=0$ it will be necessary to
make the stronger assumption $V\in L^{1}_{2}$.
In this case, we know that the Jost functions are $C^{1}$
in both variables and we shall study the behaviour of the
functions
\begin{equation}\label{eq.C}
    C_\pm(\xi,x)=
    \int_{\R}e^{-2i\lambda\xi}
    \partial_\lambda m_\pm
    (\lambda,x)d\xi
    \equiv 2i\xi B_{\pm}(\xi,x).
\end{equation}
As above, a direct application of the Deift-Trubowitz
estimate gives an optimal bound only on a half line. Indeed,
if we multiply \eqref{eq.DTbase} by $2\xi$ and integrate in
$\xi$ we obtain
\begin{equation}\label{eq.1C}
    \|C_{+}(\cdot,x)\|_{L^{1}}\leq
        2e^{\gamma(x)}
        \int_{0}^{\infty}\xi\int_{x+\xi}^{\infty}|V(t)|dtd\xi
\end{equation}
and after exchanging the order of integration we get
\begin{equation}\label{eq.2C}
    \|C_{+}(\cdot,x)\|_{L^{1}}\leq
        2e^{\gamma(x)}
        \int_{x}^{\infty}(t-x)^{2}|V(t)|dt.
\end{equation}
Notice that
\begin{equation}\label{eq.stgamma}
    \gamma(x)\leq \|V\|_{L^{1}_{1}}
        \quad\text{for ${x}\geq0$},\qquad
    \gamma(x)\leq \|V\|_{L^{1}_{1}}+|x|\cdot\|V\|_{L^{1}}
        \quad\text{for ${x}\leq0$};
\end{equation}
thus we obtain
\begin{equation}\label{eq.3C}
    \|C_{+}(\cdot,x)\|_{L^{1}}\leq C(\|V\|_{L^{1}_{2}})
        \quad\text{for ${x}\geq0$}
\end{equation}
but we can only get an exponential growth
for negative $x$ (similar computations
for $C_{-}$).

We can improve this estimate by a different argument:

\begin{lem}\label{lem.C}
    Assume $V\in L^1_2$; then the functions
    $C_{\pm}(\xi,x)=2i\xi B_{\pm}(\xi,x)$ satisfy
    \begin{equation}\label{eq.stimaC}
        \|C_\pm(\cdot,x)\|_{L^1}
        \leq C
        \quad\text{for $\pm{x}\geq0$},
        \qquad
        \|C_\pm(\cdot,x)\|_{L^1}
        \leq C\langle x
        \rangle^{2}
        \quad\text{for $\pm{x}\leq0$}
    \end{equation}
    for some constant $C$ depending on $\|V\|_{L^{1}_{2}}$.
\end{lem}

\begin{proof}
We will consider only $C_+$; the
proof for $C_-$ is identical. We have already estimated
$C_{+}$ on the positive half-line in \eqref{eq.3C}.
To prove the estimate for $x<0$ we start again from
Marchenko's equation \eqref{eq.marchenko}; if we multiply
both sides by $2\xi$ and integrate in $\xi$ we obtain
\begin{equation}\label{eq.marchC}
    \|C_+(\cdot,x)\|_{L^{1}}\leq
    2\int_{0}^{\infty}\int_{x+\xi}^{\infty}|V(t)|\cdot \xi \,dtd\xi
    +2\int_{0}^{\infty}d\xi
         \int_0^\xi\,d\sigma
         \int_{x+\xi-\sigma}^{\infty}
         |V(t)|\cdot|B_+(\sigma,t)|\cdot\xi\,dt.
\end{equation}
The first term gives, after exchanging the order of integration,
\begin{equation}\label{eq.4C}
    2\int_{0}^{\infty}\int_{x+\xi}^{\infty}|V(t)|\cdot \xi \,dtd\xi
    =
    2\int_{x}^{\infty}(t-x)^{2}|V(t)|dt.
\end{equation}
Moreover, if we remark that
\begin{equation*}
    \int_{0}^{\infty}d\xi
         \int_0^\xi\,d\sigma
         \int_{x+\xi-\sigma}^{\infty}F\, dt=
         \int_{0}^{\infty}d\sigma
         \int_x^\infty\,dt
         \int_{\sigma}^{\sigma+t-x}F\, d\xi
\end{equation*}
we see that the last integral in \eqref{eq.marchC} is equal to
\begin{equation}\label{eq.5C}
        \int_{0}^{\infty}d\sigma
         \int_x^\infty
         |V(t)|\cdot|B_+(\sigma,t)|\cdot
         [(t-x)^{2}+2\sigma(t-x)]
         \,dt
\end{equation}
and hence by \eqref{eq.marchC}, \eqref{eq.4C}, \eqref{eq.5C}
we get
\begin{equation}\label{eq.6C}
    \|C_+(\cdot,x)\|_{L^{1}}\leq
    \Phi(x)
\end{equation}
where
\begin{equation*}
    \Phi(x)\equiv
       2\int_{x}^{\infty}(t-x)^{2}|V(t)|dt+
          \int_x^\infty
         |V(t)|\cdot\left[
         (t-x)^{2}\|B_+(\cdot,t)\|_{L^{1}}+
         (t-x)\|C_+(\cdot,t)\|_{L^{1}}
         \right]dt.
\end{equation*}
In particular we see that for all $x$
\begin{equation*}
    \Phi(x)\geq0,\qquad
    \Phi'(x)\leq0,\qquad
    \Phi''(x)\geq0
\end{equation*}
i.e., $\Phi$ is a nonnegative, decreasing and convex function
on $\R$. By differentiating twice $\Phi$ we obtain
\begin{eqnarray}
    \Phi''(x) & = &
      \int_{x}^{\infty}|V(t)|\Bigl(4+2
       \|B_+(\cdot,t)\|_{L^{1}}\Bigr)dt +
    |V(x)|\cdot
       \|C_+(\cdot,x)\|_{L^{1}}\nonumber
       \\ \  & \leq &
    C_{0}(\|V\|_{L^{1}_{1}})+
    |V(x)|\cdot
       \|C_+(\cdot,x)\|_{L^{1}}\label{eq.phi1}
\end{eqnarray}
where we have used the estimate \eqref{eq.stimaB}
already proved for $B_{+}$. By \eqref{eq.6C} we obtain that
$\Phi(x)$ satisfies the differential inequality
\begin{equation}\label{eq.phi2}
    \Phi''(x)\leq
       C_{0}(\|V\|_{L^{1}_{1}})+
       |V(x)|\cdot\Phi(x).
\end{equation}
To make the following argument more clear we apply the simple
change of variables
\begin{equation*}
    \Psi(x)\equiv\Phi(-x)
\end{equation*}
so that the function $\Psi$ is nonnegative, convex and increasing:
\begin{equation*}
    \Psi(x)\geq0,\qquad
    \Psi'(x)\geq0,\qquad
    \Psi''(x)\geq0
\end{equation*}
and satisfies the inequality
\begin{equation}\label{eq.psi2}
    \Psi''(x)\leq
       C_{0}(\|V\|_{L^{1}_{1}})+
       |V(-x)|\cdot\Psi(x).
\end{equation}
Integrating \eqref{eq.psi2} from 0 to $x>0$ we obtain
\begin{equation*}
    \Psi'(x)\leq\Psi'(0)+ C_{0}x+
       \int_{0}^{x}|V(-s)|\cdot\Psi(s)ds;
\end{equation*}
writing
\begin{equation*}
    \int_{0}^{x}|V(-s)|\cdot\Psi(s)ds\equiv
    \int_{0}^{x}|V(-s)|\cdot\int_{0}^{s}\Psi'(t)dtds+
                \int_{0}^{x}|V(-s)|\cdot\Psi(0)ds
\end{equation*}
and noticing that $\Psi'(t)$ is increasing and positive, we obtain
\begin{equation*}
    \int_{0}^{x}|V(-s)|\cdot\Psi(s)ds\leq
    \int_{0}^{x}|V(-s)|\cdot s\cdot\Psi'(s)ds+
                \Psi(0)\int_{0}^{x}|V(-s)| ds
\end{equation*}
In conclusion we have proved that
\begin{equation}\label{eq.psi3}
    \Psi'(x)\leq C_{0}x+C_{1}\Psi(0)+
     \int_{0}^{x}|V(-s)|\cdot s\cdot\Psi'(s)ds
\end{equation}
with $C_{0},C_{1}$ depending on $\|V\|_{L^{1}_{1}}$.

We need now the following version of Gronwall's Lemma: if
\begin{equation*}
    \phi(x)\leq a+bx+\int_{0}^{x}\alpha(s)\phi(s)ds,\qquad
         x\geq0
\end{equation*}
for some constants $a,b$ and some $L^{1}$ function
$\alpha(s)\geq0$, then
\begin{equation}\label{eq.gron}
    \phi(x)\leq C(\|\alpha\|_{L^{1}})\cdot (a+bx) ,\qquad
       x\geq0.
\end{equation}
To check \eqref{eq.gron}, just consider the auxiliary function
\begin{equation*}
    \psi(x)=\left(a+bx+\int_{0}^{x}\alpha(s)\phi(s)ds\right)
           e^{-\int_{0}^{x}\alpha(s)ds}
\end{equation*}
which has the property
\begin{equation*}
    \psi'(x)\equiv
       \left(b+\alpha(x)\left(\phi(x)-a-bx-\int_{0}^{x}\alpha\phi ds
       \right)\right) e^{-\int_{0}^{x}\alpha(s)ds}\leq
       b e^{-\int_{0}^{x}\alpha(s)ds}\leq b;
\end{equation*}
then it is clear that
\begin{equation*}
    \psi(x)\leq\psi(0)+bx\equiv a+bx
\end{equation*}
and recalling that $\phi(x)\leq\psi(x)e^{\int_{0}^{x}\alpha(s)ds}$
we obtain \eqref{eq.gron}.

Thus if we apply \eqref{eq.gron} to
\eqref{eq.psi3} with the choice $\phi\equiv\Psi'$, we obtain
\begin{equation}\label{eq.psi4}
    \Psi'(x)\leq
      C_{2}(\|V\|_{L^{1}_{1}})\Bigl[
         |x|+\Psi(0)+\Psi'(0)\Bigr]
\end{equation}
and integrating one last time from $0$ to $x>0$
we arrive at
\begin{equation}\label{eq.psi5}
    \Psi(x)\leq
      \Bigl[C_{3}(\|V\|_{L^{1}_{1}})+
         \Psi(0)+\Psi'(0)\Bigr]\cdot\xx^{2},\qquad x\geq0.
\end{equation}
To conclude the proof it is sufficient to estimate $\Psi(0)$
and $\Psi'(0)$. Indeed, using  the preceding estimates for $B_{+}$
and $C_{+}$ on the positive half line we have immediately
\begin{equation*}
    \Psi(0)\equiv\Phi(0)\equiv
           \int_0^\infty
         |V(t)|\cdot\left[4t^{2}+2t^{2}
         \|B_+(\cdot,t)\|_{L^{1}}+t
        |C_+(\cdot,t)\|_{L^{1}}
         \right]dt\leq C(\|V\|_{L^{1}_{2}})
\end{equation*}
and
\begin{equation*}
    \Psi'(0)\equiv-\Phi'(0)\equiv
           \int_0^\infty
         |V(t)|\cdot\left[4t+2t
         \|B_+(\cdot,t)\|_{L^{1}}+
        |C_+(\cdot,t)\|_{L^{1}}
         \right]dt\leq C(\|V\|_{L^{1}_{2}}).
\end{equation*}
Recalling \eqref{eq.6C}, we conclude that
\begin{equation*}
    \|C_{+}(\cdot,x)\|_{L^{1}}\leq\Phi(x)=\Psi(-x)\leq
        C(\|V\|_{L^{1}_{2}})\xx^{2},\qquad
        x\leq0
\end{equation*}
as claimed.
\end{proof}

A useful consequence of \eqref{eq.stimaC} is an estimate of the
Fourier transform of the functions
\begin{equation}\label{eq.incr}
    n_{\pm}(\lambda,x)=
     \frac{m_{\pm}(\lambda,x)-m_{\pm}(0,x)}{\lambda}
\end{equation}
which are clearly related to the derivatives
$\partial_{\lambda}m_{\pm}$; the usefulness of these
quantities in the resonant case had already been remarked
in \cite{ArbYaj00}.

\begin{cor}\label{cor.n}
    Assume $V\in L^1_2$; then the functions
    $\widetilde C_{\pm}(\xi,x)=
        \int_{\R}e^{-2i\lambda\xi}n_{\pm}(\lambda,x)d\lambda$
    satisfy
    \begin{equation}\label{eq.stimaC2}
        \| \widetilde C_\pm(\cdot,x)\|_{L^1}
        \leq C
        \quad\text{for $\pm{x}\geq0$},
        \qquad
        \| \widetilde C_\pm(\cdot,x)\|_{L^1}
        \leq C\langle x
        \rangle^{2}
        \quad\text{for $\pm{x}\leq0$}
    \end{equation}
    for some constant $C$ depending on $\|V\|_{L^{1}_{2}}$.
\end{cor}

\begin{proof}
We can write
\begin{equation*}
    n_{\pm}(\lambda,x)=
     \frac{m_{\pm}(\lambda,x)-m_{\pm}(0,x)}{\lambda}=
      \int_{0}^{1}\partial_{\lambda}
          m_{\pm}(\lambda s,x)ds
\end{equation*}
and this implies, by Fubini's theorem and the rescaling
properties of the one dimensional Fourier transform,
\begin{equation*}
    \widetilde C_\pm(\xi,x)=
        \int_{0}^{1}\mathcal F_{\lambda\to\xi}
            \left( \partial_{\lambda}
             m_{\pm}(\lambda s,x) \right)ds=
        \int_{0}^{1}s^{-1}C_{\pm}(\xi/s,x)ds.
\end{equation*}
The integral Minkowski inequality now gives
\begin{equation*}
    \| \widetilde C_\pm(\cdot,x)\|_{L^1}\leq
    \int_{0}^{1}s^{-1}
         \|C_\pm(\cdot/s,x)\|_{L^1}ds\equiv
    \int_{0}^{1}
         \|C_\pm(\cdot,x)\|_{L^1}ds\equiv
         \|C_\pm(\cdot,x)\|_{L^1}
\end{equation*}
and by \eqref{eq.stimaC} the proof is concluded.
\end{proof}

We conclude this section by studying the Fourier properties
of the Wronskian $W(\lambda)$
\begin{equation*}
    W(\lambda)=
    f_{+}(\lambda,0)\partial_{x}f_{-}(\lambda,0)-
       \partial_{x}f_{+}(\lambda,0)f_{-}(\lambda,0)
\end{equation*}
which can be equivalently written
\begin{equation*}
    W(\lambda)=
      m_{+}(\lambda,0)\partial_{x}m_{-}(\lambda,0)-
      \partial_{x}m_{+}(\lambda,0)m_{-}(\lambda,0)-
      2i\lambda
          m_{+}(\lambda,0)m_{-}(\lambda,0).
\end{equation*}
Notice that the following result is also proved in
\cite{GoldbergSchlag04} by partly different arguments.

\begin{lem}\label{lem.W}
    Let $\chi(\lambda)\in C^{\infty}_{0}(\R)$ be
    a smooth cutoff. If $V\in L^{1}_{1}(\R)$
    and $W(0)\neq0$ then
    \begin{equation}\label{eq.FW}
         \mathcal F\left(\frac{\chi(\lambda)}{W(\lambda)}\right)\in
         L^{1}(\R).
    \end{equation}
    On the other hand, if $V\in L^{1}_{2}(\R)$
    and $W(0)=0$ then
    \begin{equation}\label{eq.FWl}
         \mathcal F\left(\frac{\chi(\lambda)\lambda}{W(\lambda)}\right)\in
         L^{1}(\R).
    \end{equation}
\end{lem}

\begin{proof}
Let $\chi_{1}\in C^{\infty}_{0}(\R)$ be a second cutoff such that
$\chi_{1}\equiv 1$ on the support of $\chi$.
By the Deift-Trubowitz estimates
(see Lemma \ref{lem.lemma3DT})
we know that both $m_{\pm}(\lambda,0)-1$ and
$\partial_{x}m_{\pm}(\lambda,0)$ have Fourier transform
in $L^{1}$; then writing
\begin{eqnarray*}
    \chi_{1}W(\lambda) & \equiv &
      \chi_{1}(\lambda)
      m_{+}(\lambda,0)\partial_{x}m_{-}(\lambda,0)-
     \chi_{1}(\lambda)
      \partial_{x}m_{+}(\lambda,0)m_{-}(\lambda,0)
      \\ \  & \  & -
      2i\lambda
      \chi_{1}(\lambda) m_{+}(\lambda,0)m_{-}(\lambda,0)
\end{eqnarray*}
we see that $\chi_{1}W$ can be written as a sum of products
in which each factor has a Fourier transform in $L^{1}$, and we
conclude that $\chi_{1}W$ has Fourier transform in $L^{1}$.

Recall now that by Wiener's Lemma, if a function $a(\lambda)$ does not
vanish on the support of $b(\lambda)$ and both
$\hat a,\hat b\in L^{1}$, we have also
$\mathcal F(b/a)\in{L^{1}}$.
This implies that
\begin{equation*}
    \left\|\mathcal F_{\lambda\to\xi}
    \left(\frac{\chi(\lambda)}
    {{W(\lambda)}}
    \right)\right\|_{L^1_\xi}\equiv
    \left\|\mathcal F_{\lambda\to\xi}
    \left(\frac{\chi(\lambda)}
    {\chi_{1}(\lambda){W(\lambda)}}
    \right)\right\|_{L^1_\xi}<\infty.
\end{equation*}

Consider now the resonant case with $V\in L^{1}_{2}$. Using the
functions $n_{\pm}$ defined in \eqref{eq.incr} we can rewrite
$W$ as follows:
\begin{eqnarray*}
    W(\lambda) & = & \lambda n_{+}(\lambda,0)
             \partial_{x}m_{-}(\lambda,0)+
         \lambda m_{+}(0,0) \partial_{x}n_{-}(\lambda,0)+
         m_{+}(0,0)  \partial_{x}m_{-}(0,0)
         \\ \  & \  & +
         2i\lambda m_{+}(\lambda,0)m_{-}(\lambda,0);
\end{eqnarray*}
from this formula and the assumption $W(0)=0$
we see that the term $m_{+}(0,0)  \partial_{x}m_{-}(0,0)$
must vanish, hence we obtain
\begin{equation}\label{eq.Wsul}
    \frac{W(\lambda)}\lambda= n_{+}(\lambda,0)
             \partial_{x}m_{-}(\lambda,0)+
         m_{+}(0,0) \partial_{x}n_{-}(\lambda,0)+
         2i m_{+}(\lambda,0)m_{-}(\lambda,0).
\end{equation}
We know already that the functions
$m_{\pm}(\lambda,0)-1$,
$\partial_{x}m_{\pm}(\lambda,0)$ and $n_{\pm}(\lambda,0)$
have Fourier transform in $L^{1}$; this follows as above from
the Deift-Trubowitz estimate and from our Corollary \ref{cor.n}
(see \eqref{eq.stimaC2}). We can show that also
$\partial_{x}n_{\pm}(\lambda,0)$ have the same property; indeed,
\begin{equation*}
    n_{\pm}(\lambda,x)=
      \int_{0}^{1}\partial_{\lambda}
          m_{\pm}(\lambda s,x)ds\quad\implies\quad
         \partial_{x}n_{\pm}(\lambda,x)=
      \int_{0}^{1} \partial_{x}\partial_{\lambda}
          m_{\pm}(\lambda s,x)ds;
\end{equation*}
then by Fubini's theorem and
the rescaling properties of the Fourier transform we have
\begin{equation*}
    \mathcal F_{\lambda\to\xi}
     \left( \partial_{x}n_{\pm}(\lambda,0)\right)=
    \int_{0}^{1}2i\xi s^{-1}\partial_{x}B_{\pm}(\xi/s,0)ds
\end{equation*}
and, by the integral Minkowski inequality,
\begin{equation*}
    \|\mathcal F_{\lambda\to\xi}
     \left( \partial_{x}n_{\pm}(\lambda,x)\right)\|_{L^{1}_{\xi}}
    \leq 2\int_{0}^{1}
    \|\xi s^{-1}\partial_{x}B_{\pm}(\xi/s,0)\|_{L^{1}_{\xi}}ds=
     2\int_{0}^{1}
    \|\xi \partial_{x}B_{\pm}(\xi,0)\|_{L^{1}_{\xi}}s\,ds
\end{equation*}
and we arrive at
\begin{equation}\label{eq.dexB}
    \|\mathcal F_{\lambda\to\xi}
     \left( \partial_{x}n_{\pm}(\lambda,x)\right)\|_{L^{1}_{\xi}}
     \leq
    \|\xi \partial_{x}B_{\pm}(\xi,0)\|_{L^{1}_{\xi}}.
\end{equation}
Recalling now the Deift-Trubowitz estimate \eqref{eq.DTbase}, we
have immediately
\begin{equation*}
    |\xi \partial_{x}B_{\pm}(\xi,0)|\leq
       C|\xi|\cdot \bigl[\eta(\xi)+|V(\xi)|\bigr]
     \quad\implies\quad
    \|\xi \partial_{x}B_{\pm}(\xi,0)\|_{L^{1}_{\xi}}\leq
       C\|V\|_{L^{1}_{2}}
\end{equation*}
and this proves that the Fourier transform
of $\partial_{x}n_{\pm}(\lambda,0)$ belongs to $L^{1}(\R)$.

Now, coming back to \eqref{eq.Wsul}, and choosing a cutoff
$\chi_{1}$ as above, we see that
$\chi_{1}(\lambda)W(\lambda)/\lambda$ can be written as a sum of
products of functions with Fourier transform in $L^{1}$ and hence it
also has Fourier transform in $L^{1}$; applying Wiener's Lemma
exactly as before we conclude the proof.
\end{proof}

\section{The low energy analysis}\label{sec.bassenerg}

In this section we shall study the low energy part of the wave operator
$W_{+}$; the estimate for $W_{-}$ is completely analogous.
By the stationary representation formula
(see e.g. \cite{Yajima95-waveopN}), given a cutoff
$\Phi(\lambda^{2})$ supported
near zero, we can represent the low energy part
of $W_{+}$ as follows:
\begin{equation}\label{eq.WRV}
    W_+\Phi(H_{0})g=\Phi(H_{0})g-
    \frac{1}{\pi i}\int_{0}^{+\infty}
       R_V(\lambda^{2}-i0)V\,\,
             \Im R_0(\lambda^{2}+i0)
          \lambda\Phi(\lambda^{2})g\,d\lambda.
\end{equation}
Thus it is sufficient to study the boundedness
in $L^{p}$ of the operator
\begin{equation}\label{eq.A}
    Ag:=
    \int_{0}^{+\infty}
    R_V(\lambda^2-i0)V\,
    \Im R_0(\lambda^2+i0)\lambda\chi(\lambda)g
    \,d\lambda
\end{equation}
for an even cutoff function
$\chi(\lambda)=\Phi_{0}(\lambda^{2})\in C_{0}^{\infty}(\R)$.

As remarked in the Introduction, an $L^{\infty}-L^{\infty}$
estimate will be impossible in general, owing to the presence of
a Hilbert transform term in the wave operator. We recall
that the \emph{Hilbert transform} on $\R$ is the operator
\begin{equation*}
    \HH g(y)=\frac1\pi
      V.P.\int_{\R}\frac{g(s)}{y-s}ds,
\end{equation*}
which can be equivalently defined as a Fourier multiplier with symbol
$(2\pi i)^{-1}\mathop{\mathrm{sgn}}\xi$:
\begin{equation*}
    \HH g=
     \frac1{2\pi i}\int e^{iy\lambda}\frac{\lambda}{|\lambda|}
     \widehat g(\lambda)d\lambda.
\end{equation*}
We also recall that $\HH^{2}=-1$, and that
$\HH$ is a bounded operator on $L^{p}$
for all $1<p<\infty$, but not on $L^{1}$ and on $L^{\infty}$.

In order to state a simple but useful interpolation lemma
we introduce the Banach spaces
\begin{equation}\label{eq.Lp0}
    L^{\infty}_{0}
    =\{g\in L^{\infty}\colon
       g\to0\text{ as }|x|\to\infty\},\qquad
       \|g\|_{L^{\infty}_{0}}=\| g\|_{L^{\infty}}.
\end{equation}
and
\begin{equation}\label{eq.LH}
    \LpH=\{g\in L^{p}\colon
           \HH g\in L^{p}\},\qquad
       \|g\|_{\LpH}=\| g\|_{L^{p}}+\|\HH g\|_{L^{p}}.
\end{equation}
Notice that the last definition
is relevant only when $p=1$ or $p=\infty$,
since we have otherwise
\begin{equation}\label{eq.LH2}
    \LpH\simeq L^{p}\quad\text{for}\quad 1<p<\infty.
\end{equation}
Our interpolation lemma is then the following:

\begin{lem}\label{lem.int}
Let $T$ be a bounded operator on $L^{2}$, and assume that
\begin{equation}\label{eq.int1}
    \|Tg\|_{L^{\infty}}+ \|T^{*}g\|_{L^{\infty}}\leq
     C\|g\|_{L^{\infty}}+
     C\|\HH g\|_{L^{\infty}},\qquad\forall
     g\in L^{\infty}\cap\LiH\cap L^{1}.
\end{equation}
Then $T$ and $T^{*}$ can be extended to bounded operators
on $L^{p}$ for all $1<p<\infty$.
\end{lem}

\begin{proof}
From the standard theory of complex interpolation it is
known that
\begin{equation}\label{eq.int0}
    [L^{p_{0}},L^{p_{1}}]_{\theta}=L^{p_{\theta}},\qquad
     \frac1{p_{\theta}}=\frac{1-\theta}{p_{0}}+\frac\theta{p_{1}},
     \quad 0<\theta<1,\quad 1\leq p_{0}<p_{1}\leq\infty
\end{equation}
and it is also well known that $L^{\infty}$ can be replaced by
$L^{\infty}_{0}$ in the interpolation scale:
\begin{equation}\label{eq.int2}
    [L^{p},L^{\infty}_{0}]_{\theta}=L^{p_{\theta}},\qquad
     \frac1{p_{\theta}}=\frac{1-\theta}{p},
     \quad 0<\theta<1,\quad 1\leq p<\infty
\end{equation}
(see \cite{BL}). By an elementary modification in the
construction, keeping into account
\eqref{eq.LH2}, it is easy to see that we have also
\begin{equation}\label{eq.int3}
    [L^{p},L^{\infty}_{0}\cap\LiH]_{\theta}=L^{p_{\theta}},\qquad
     \frac1{p_{\theta}}=\frac{1-\theta}{p},
     \quad 0<\theta<1,\quad 1< p<\infty
\end{equation}
where $L^{\infty}_{0}\cap\LiH$ is endowed with the norm
$\|g\|_{L^{\infty}\cap\LiH}=\|g\|_{L^{\infty}}+\|g\|_{\LiH}$.
Now, by a density argument we see that \eqref{eq.int1} implies
that $T,T^{*}$ can be extended to bounded operators from
$L^{\infty}_{0}\cap\LiH$ to $L^{\infty}$, and on the other hand
they are bounded on $L^{2}$ by assumption. Using \eqref{eq.int3},
by interpolation we obtain that $T,T^{*}$ are bounded on
all $L^{p}$ for $2\leq p<\infty$, and by duality we conclude the
proof.
\end{proof}

\begin{remark}\label{rem.int}
In the endpoint case $p=\infty$ we can modestly improve
\eqref{eq.int1} to
\begin{equation}\label{eq.int4}
    \|Tg\|_{L^{\infty}}\leq
     C\|g\|_{L^{\infty}}+
     C\|\HH g\|_{L^{\infty}},\qquad\forall
     g\in L^{\infty}\cap\LiH\cap L^{p}
\end{equation}
for some $p<\infty$; this follows immediately by a density argument.
Moreover, in the opposite endpoint $p=1$, by duality, we obtain that
\begin{equation}\label{eq.L1}
    \|Tg\|_{L^{1}+\LuH}\leq C\|g\|_{L^{1}}
\end{equation}
where $L^{1}+\LuH$ is the Banach space with norm
\begin{equation*}
    \|g\|_{L^{1}+\LuH}\equiv
        \inf\{\|g_{1}\|_{L^{1}}+\|g_{2}\|_{\LuH}\colon
          g=g_{1}+{g_{2}},\ g_{1}\in L^{1},g_{2}\in\LuH\}.
\end{equation*}
\end{remark}

We are now ready to prove our estimate of the low frequency part
of the wave operator:

\begin{lem}\label{lem.nonres}
    Assume $V\in L^1_1$ and the nonresonant condition
    $W(0)\neq0$ is satisfied. Let $\Phi(\lambda^{2})$ be
    a smooth compactly supported cutoff function.
    Then the low energy parts of the wave operators
    $W_{\pm}$ satisfy the estimates
    \begin{equation}\label{eq.stimaA1}
        \|W_{\pm}\Phi(H_{0})g\|_{L^\infty}\leq
        C\left(\|g\|_{L^\infty}+\|\HH g\|_{L^{\infty}}\right)
        \qquad\forall g\in L^{1}\cap L^{\infty}\cap\LiH
    \end{equation}
    and hence can be extended to bounded operatos on $L^{p}$,
    for all $1<p<\infty$. The same properties hold for the
    conjugate operators $\Phi(H_{0})W^{*}_{\pm}$.
\end{lem}

\begin{proof}
The proof for the operators $W_{\pm}$ and $W^{*}_{\pm}$
is completely analogous, hence we shall focus on the
estimate for $W_{+}$. By Lemma \ref{lem.int}, it
is sufficient to prove that $W_{+}\Phi(H_{0})$
satisfies \eqref{eq.stimaA1}; morevoer,
using the stationary representation formula \eqref{eq.WRV},
we are reduced to estimate the operator $A$
defined by \eqref{eq.A}.

By the explicit expression of the kernel of $R_{V}$ in terms
of the Jost functions \eqref{eq.jostRV},
we can split $A$ as $A=A_1+A_2$ where
(forgetting constants)
\begin{equation}\label{eq.A1}
    A_1g(x)=\int_{0}^{+\infty}
    \,d\lambda\int_{x<y}
    \,dy
    \frac{f_+(-\lambda,y)
    f_-(-\lambda,x)}{W(-\lambda)}V(y)
    \lambda\chi(\lambda)\Im R_{0}(\lambda^{2}+i0)g(y)
\end{equation}
and
\begin{equation}\label{eq.A2}
    A_2g(x)=\int_{0}^{+\infty}
    \,d\lambda\int_{x<y}
    \,dy
    \frac{f_+(-\lambda,x)
    f_-(-\lambda,y)}{W(-\lambda)}V(y)
    \lambda\chi(\lambda)\Im R_{0}(\lambda^{2}+i0)g(y).
\end{equation}
Using the relations \eqref{eq.m} and
\begin{equation*}
    m_\pm(-\lambda,x)=
    \overline{m_\pm(\lambda,x)}, \qquad
    W(-\lambda)=
    \overline{W(\lambda)}
\end{equation*}
(see. e.g. \cite{DeiTru}),
we have
\begin{equation}\label{eq.A1m}
    A_1g(x)=\int_{0}^{+\infty}
    \,d\lambda\int_{x<y}\,dy
    \frac{\overline{m_+(\lambda,y)}
   \overline{m_-(\lambda,x)}}{\overline{W(\lambda)}}V(y)
    e^{i\lambda(x-y)}
    \lambda\chi(\lambda)\Im R_{0}(\lambda^{2}+i0)g(y)
\end{equation}
\begin{equation}\label{eq.A2m}
    A_2g(x)=\int_{0}^{+\infty}
    \,d\lambda\int_{x>y}\,dy
    \frac{\overline{m_+(\lambda,x)}
   \overline{m_-(\lambda,y)}}{\overline{W(\lambda)}}V(y)
    e^{i\lambda(y-x)}
    \lambda\chi(\lambda)\Im R_{0}(\lambda^{2}+i0)g(y).
\end{equation}

In the following we shall estimate the operator $A_{1}$;
the proof for $A_{2}$ is completely analogous.
By Fubini's Theorem we can exchange the order of integration
and rewrite \eqref{eq.A1m} as follows:
\begin{equation}\label{eq.A1F}
    A_1g(x)=\int_{x<y}
    \mathcal F_{\lambda\to\xi}
    \left.\left(
    \frac{\overline{m_+(\lambda,y)m_-(\lambda,x)}}
         {\overline{W(\lambda)}}
          \chi(\lambda)
          \uno_{(0,+\infty)}(\lambda)
          \Im R_{0}(\lambda^{2}+i0)g(y)
      \right)\right\vert
       _{\xi=x-y}V(y)dy
\end{equation}
where $\mathcal F$ denotes the standard Fourier transform from the
$\lambda$ to the $\xi$
variable and $\uno_{(0,+\infty)}$ is the characteristic function of
the half line $(0,+\infty)$.

Now choose a $C^\infty_0$
cutoff function $\psi(\lambda)$ such that
$\psi\equiv1$ on $\mathop{\mathrm{supp}}\chi$ ;
then the function
\begin{equation*}
    G(\lambda ,x,y)=
    \frac{\overline{m_+(\lambda,y)m_-(\lambda,x)}}
         {\overline{W(\lambda)}}
          \chi(\lambda)
          \uno_{(0,+\infty)}(\lambda)
          \Im R_{0}(\lambda^{2}+i0)g(y)
\end{equation*}
can be written as a product
\begin{equation}\label{eq.G}
    G(\lambda ,x,y)=
         F_{1}(\lambda,y)F_{2}(\lambda,x)F_{3}(\lambda)
          F_{4}(\lambda,y)
\end{equation}
where
\begin{equation*}
    F_{1}(\lambda,y)=\overline{m_+(\lambda,y)}
         \psi(\lambda),\qquad
    F_{2}(\lambda,x)=\overline{m_-(\lambda,y)}
        \psi(\lambda),\qquad
    F_{3}(\lambda)=
       \frac{\psi(\lambda)}
         {(\lambda)\overline{W(\lambda)}}
\end{equation*}
and
\begin{equation*}
    F_{4}(\lambda,y)=
          \chi(\lambda)
          \uno_{(0,+\infty)}(\lambda)
          \Im R_{0}(\lambda^{2}+i0)g(y).
\end{equation*}
We are interested in the Fourier transform of $G$ with respect
to $\lambda$; this can be written as the convolution of the
transforms $\widehat F_{j}$, $j=1,2,3,4$.

By Lemma \ref{lem.W} (see \eqref{eq.FW}) we
already know that
\begin{equation}\label{eq.F3}
    \|\widehat F_{3}(\xi)\|_{L^{1}_{\xi}}=C_{0}<\infty.
\end{equation}

Consider now $\widehat F_{1}(\xi,y)$, which can be written
\begin{equation*}
    \widehat F_{1}(\xi,y)=
     \mathcal F((\overline{m_{+}(\lambda,y)}-1)\psi_{1}
         +\psi_{1})=
     \overline{B_{+}(-\xi/2,y)}*\widehat\psi_{1}+\widehat\psi_{1}
\end{equation*}
(the inessential factor $1/2$ comes from the nonstandard
Fourier transform used in Definition \eqref{eq.B}, and
the minus sign from the conjugation).
Recalling  Lemma \ref{lem.B}, we get
\begin{equation*}
    \|\widehat F_{1}(\cdot,y)\|_{L^{1}}
        \leq
     \begin{cases}
              C &\text{for ${y}\geq0$}\\
           C\langle y\rangle&\text{for ${y}\leq0$}
     \end{cases}
        \qquad
\end{equation*}
for some $C$ depending on $\|V\|_{L^{1}_{1}}$. The same
argument gives
\begin{equation*}
    \|\widehat F_{2}(\cdot,x)\|_{L^{1}}
        \leq
     \begin{cases}
              C &\text{for ${x}\leq0$}\\
           C\langle x\rangle&\text{for ${x}\geq0$}
     \end{cases}
        \qquad
\end{equation*}
with the same $C$. Recalling that in \eqref{eq.A1F}
we have $x<y$, we can write
\begin{equation*}
    \|\mathcal F(F_{1}F_{2})\|_{L^{1}_{\xi}}\leq
    \|\widehat F_{1}(\cdot,x)\|_{L^{1}}
    \|\widehat F_{2}(\cdot,y)\|_{L^{1}}\leq
    \begin{cases}
        C\langle y\rangle&\text{for $x<y<0$}\\
        C &\text{for $x<0<y$}\\
        C\langle x\rangle\leq C\langle y\rangle&\text{for $0<x<y$}
    \end{cases}
\end{equation*}
and in conclusion
\begin{equation}\label{eq.F1F2}
    \|\mathcal F(F_{1}F_{2})\|_{L^{1}_{\xi}}\leq
               C(\|V\|_{L^{1}_{1}})\cdot\langle y\rangle.
\end{equation}

Coming back to $G(\lambda,x,y)$,
if we put together \eqref{eq.F3} and \eqref{eq.F1F2}
and we use Young's inequality,
we have proved that, for $x<y$,
\begin{equation}\label{eq.GF}
    \|\widehat G(\cdot,x,y)\|_{L^{\infty}}\leq
       C(\|V\|_{L^{1}_{1}})\cdot \langle y\rangle\cdot
       \|\widehat F_{4}(\cdot,y)\|_{L^{\infty}}.
\end{equation}
It remains to estimate
\begin{equation*}
       \|\widehat F_{4}(\cdot,y)\|_{L^{\infty}}\equiv
     \sup_{\xi}
       \left|\mathcal F_{\lambda\to\xi}\left(
           \chi(\lambda)\uno_{(0,+\infty)}(\lambda) \lambda
           \Im R_{0}(\lambda^{2}+i0)g(y)
           \right)\right|.
\end{equation*}
We have
\begin{eqnarray*}
    \mathcal F\left(
           \chi(\lambda)\uno_{(0,+\infty)}(\lambda) \lambda
           \Im R_{0}(\lambda^{2}+i0)g(y)
           \right) & = &
     \int_{0}^{\infty}
         e^{i\lambda\xi}\lambda\chi(\lambda)
                  \Im R_{0}(\lambda^{2}+i0)g(y)d\lambda
        \\ \  & \equiv &
        C e^{i\xi\sqrt{H_{0}}}\chi(\sqrt{H_{0}})g
\end{eqnarray*}
by the spectral theorem. Now we remark that the function
\begin{equation*}
    U(\xi,y)=e^{i\xi\sqrt{H_{0}}}\chi(\sqrt{H_{0}})g
\end{equation*}
is a solution of the one dimensional wave equation
\begin{equation*}
    U_{\xi\xi}+H_{0}U\equiv
    U_{\xi\xi}-U_{yy}=0,
\end{equation*}
with initial data
\begin{equation*}
    U(0,y)=U_{0}(y)=\chi(\sqrt{H_{0}})g,\qquad
    U_{\xi}(0,y)=U_{1}(y)=i\sqrt{H_{0}}\chi(\sqrt{H_{0}})g.
\end{equation*}
By the explicit representation formula of the solution to the
wave equation we have then
\begin{equation*}
    U(\xi,y)=\frac{U_{0}(\xi+y)+U_{1}(\xi-y)}2
         +\frac12\int_{\xi-y}^{\xi+y}U_{1}(\sigma)d\sigma.
\end{equation*}
The first term is easy to bound:
\begin{equation}\label{eq.u0}
    \left|\frac{U_{0}(\xi+y)+U_{1}(\xi-y)}2\right|
    \leq \|U_{0}\|_{L^{\infty}}=
    \|\chi(\sqrt{H_{0}})g\|_{L^\infty}\leq
    C\|g\|_{L^\infty}
\end{equation}
since $\chi(\sqrt{H_{0}})$ is bounded on $L^{\infty}$ as
it is well known. On the other hand, we can write
\begin{equation*}
    U_{1}(y)=i\sqrt{H_{0}}\chi(\sqrt{H_{0}})g(y)=
    \int e^{i\lambda y}{|\lambda|}
        \chi(\lambda)\widehat g(\lambda)d\lambda\equiv
    \int e^{i\lambda y}(-i\lambda)
        \chi(\lambda)
        \frac{i\lambda}{|\lambda|}
        \widehat g(\lambda)d\lambda
\end{equation*}
and this implies, apart from a constant,
\begin{equation*}
    U_{1}(y)=\frac{d}{dy}\chi(\sqrt{H_{0}}) \HH g
    \ \ \implies\ \
   \int_{\xi-y}^{\xi+y}U_{1}(\sigma)d\sigma=
   \chi(\sqrt{H_{0}}) \HH g(\xi+y)-
   \chi(\sqrt{H_{0}}) \HH g(\xi-y).
\end{equation*}
In conclusion
\begin{equation*}
    \left|\frac12\int_{\xi-y}^{\xi+y}U_{1}(\sigma)d\sigma\right|\leq
    \|\chi(\sqrt{H_{0}}) \HH g\|_{L^{\infty}}\leq
    C\|\HH g\|_{L^{\infty}}
\end{equation*}
and summing up we have proved that
\begin{equation}\label{eq.F4}
    \|\widehat F_{4}\|_{L^{\infty}_{\xi,y}}\leq
   C\|U\|_{L^{\infty}_{\xi,y}}\leq
   C\left(\|g\|_{L^\infty}+\|\HH g\|_{L^{\infty}}\right).
\end{equation}
By \eqref{eq.A1F}, \eqref{eq.GF} and \eqref{eq.F4} we
finally obtain
\begin{equation*}
    \|A_{1}g\|_{L^{\infty}}\leq
      C(\|V\|_{L^{1}_{1}})\cdot
      \left(\|g\|_{L^\infty}+\|\HH g\|_{L^{\infty}}\right).
\end{equation*}
The operator $A_{2}$ can be estimated in a similar way,
and this concludes the proof of the Lemma.
\end{proof}

We pass to the analysis of the resonant case $W(0)=0$.

%
%

\begin{lem}\label{lem.res}
    Assume $V\in L^1_2$ and we are in the resonant case
    $W(0)=0$. Let $\Phi(\lambda^{2})$ be
    a smooth compactly supported cutoff function.
    Then the following estimate holds:
    \begin{equation}\label{eq.stimaA1bis}
        \|W_{+}\Phi(H_{0})g\|_{L^\infty}\leq
        C\left(\|g\|_{L^\infty}+\|\HH g\|_{L^{\infty}}\right)
        \qquad\forall g\in L^{1}\cap L^{\infty}\cap\LiH
    \end{equation}
    where $\HH$ is the Hilbert transform on $\R$, and
    hence can be extended to bounded operators
    on $L^{p}$, for all $1<p<\infty$. The same
    estimate holds for the
    conjugate operators $\Phi(H_{0})W^{*}_{\pm}$.
\end{lem}

\begin{proof}
As in the proof of Lemma \ref{lem.nonres}, we are reduced to
estimate the $L^{\infty}$ norm of $Ag=A_{1}g+A_{2}g$
where $A_{j}g$ are given by expressions \eqref{eq.A1m} and
\eqref{eq.A2m}. The new difficulty now is of course the denominator
$W(\lambda)$ which vanishes at $\lambda=0$.
Thus we decompose $Ag$ into several terms:
\begin{equation*}
    Ag=I_{1}+I_{2}+II_{1}+II_{2}+III_{1}+III_{2}
\end{equation*}
where, recalling the notation \eqref{eq.incr}.
\begin{equation}\label{eq.I1}
    I_{1}=\int_{0}^{+\infty}d\lambda\int_{x>y}dy
       \frac{-\lambda}{W(-\lambda)}
    {n_+(-\lambda,y)}m_-(-\lambda,x)
    e^{i\lambda(y-x)}V(y)
    \lambda\chi(\lambda)\Im R_{0}(\lambda^{2}+i0)g(y),
\end{equation}
\begin{equation}\label{eq.I2}
    I_{2}=\int_{0}^{+\infty}d\lambda\int_{x<y}dy
       \frac{-\lambda}{W(-\lambda)}
    {n_+(-\lambda,x)}m_-(-\lambda,y)
    e^{i\lambda(x-y)}V(y)
    \lambda\chi(\lambda)\Im R_{0}(\lambda^{2}+i0)g(y),
\end{equation}
\begin{equation}\label{eq.II1}
    II_{1}=\int_{0}^{+\infty}d\lambda\int_{x>y}dy
       \frac{-\lambda}{W(-\lambda)}
    {m_+(0,y)}n_-(-\lambda,x)
    e^{i\lambda(y-x)}V(y)
    \lambda\chi(\lambda)\Im R_{0}(\lambda^{2}+i0)g(y),
\end{equation}
\begin{equation}\label{eq.II2}
    II_{2}=\int_{0}^{+\infty}d\lambda\int_{x<y}dy
       \frac{-\lambda}{W(-\lambda)}
    {m_+(0,x)}n_-(-\lambda,y)
    e^{i\lambda(x-y)}V(y)
    \lambda\chi(\lambda)\Im R_{0}(\lambda^{2}+i0)g(y),
\end{equation}
\begin{equation}\label{eq.III1}
    III_{1}=\int_{0}^{+\infty}d\lambda\int_{x>y}dy
    \frac{m_+(0,y)m_{-}(0,x)}{W(-\lambda)}V(y)
    e^{i\lambda(y-x)}
    \lambda\chi(\lambda)\Im R_{0}(\lambda^{2}+i0)g(y),
\end{equation}
\begin{equation}\label{eq.III2}
    III_{2}=\int_{0}^{+\infty}d\lambda\int_{x<y}dy
    \frac{m_+(0,x)m_{-}(0,y)}{W(-\lambda)}V(y)
    e^{i\lambda(x-y)}
    \lambda\chi(\lambda)\Im R_{0}(\lambda^{2}+i0)g(y).
\end{equation}

An essential remark is the following: since $W(0)=0$,
we know that for $\lambda=0$ the Jost functions
$f_{+}(0,x)\equiv m_{+}(0,x)$
and $f_{-}(0,x)\equiv m_{-}(0,x)$ are linearly dependent, i.e.,
\begin{equation}\label{eq.dep}
    m_{-}(0,x)=c_{0}\cdot m_{+}(0,x)
\end{equation}
for some constant $c_{0}\neq0$.
Moreover, by definition $m_{\pm}(0,x)\to1$ as
$\pm x\to\infty$, and together with \eqref{eq.dep} this implies
that  $m_{\pm}(0,x)$ are bounded on $\R$:
\begin{equation}\label{eq.bdd}
    |m_{\pm}(0,x)|\leq c_{1},\qquad x\in\R.
\end{equation}
Finally, when $W(0)=0$ we have
\begin{equation}\label{eq.media}
    \int_{-\infty}^{+\infty}V(y)m_\pm(0,y)dy=0
\end{equation}
(see e.g. \cite{ArbYaj00}).

The terms of type $I$ and $II$ are handled in a way very similar
to the proof of Lemma \ref{lem.nonres}.
In order to estimate the term $I_{1}$, we write it as
\begin{equation*}
    I_{1}=\int_{x<y}
     \left.\mathcal F_{\lambda\to\xi}
     \left(G(\lambda,x,y)\right)\right|_{\xi=y-x}V(y)dy
\end{equation*}
where
\begin{equation*}
    G(\lambda,x,y)=
         F_{1}(\lambda,y)F_{2}(\lambda,x)F_{3}(\lambda)
          F_{4}(\lambda,y),
\end{equation*}
$\psi$ is a $C^\infty_0(\R)$
cutoff function such that
$\psi\equiv1$ on $\mathop{\mathrm{supp}}\chi$,
and
\begin{equation*}
    F_{1}(\lambda,y)={n_+(-\lambda,y)},\qquad
    F_{2}(\lambda,x)={m_-(-\lambda,x)}
        \psi(\lambda),\qquad
    F_{3}(\lambda)=
       \frac{\psi(\lambda)\lambda}
         {W(-\lambda)}
\end{equation*}
and
\begin{equation*}
    F_{4}(\lambda,y)=
          \chi(\lambda)
          \uno_{(0,+\infty)}(\lambda)
          \Im R_{0}(\lambda^{2}+i0)g(y).
\end{equation*}
Then we have
\begin{equation*}
    \sup_{x<y}
    \|\widehat G(\cdot,x,y)\|_{L^{\infty}}\leq
    \sup_{x<y}\left(
    \|\widehat F_{1}(\cdot,y)\|_{L^{1}}\cdot
    \|\widehat F_{2}(\cdot,x)\|_{L^{1}}\cdot
    \|\widehat F_{3}\|_{L^{1}}\cdot
    \|\widehat F_{4}(\cdot,y)\|_{L^{\infty}}
    \right)
\end{equation*}
Using Lemma \ref{lem.W} we see that
$\|\widehat F_{3}\|_{L^{1}}=C_{0}<\infty$, and
by Lemma \ref{lem.lemma3DT} and Corollary \ref{cor.n}
we obtain as before (by considering the three cases $x<y<0$,
$x<0<y$ and $0<x<y$)
\begin{equation*}
    \|\widehat F_{1}(\cdot,y)\|_{L^{1}}\cdot
    \|\widehat F_{2}(\cdot,x)\|_{L^{1}}\leq
     C(\|V\|_{L^{1}_{2}})\cdot\langle y\rangle^{2}\qquad
     \text{for $x<y$}.
\end{equation*}
Thus we arrive at
\begin{equation*}
    \|\widehat G(\cdot,x,y)\|_{L^{\infty}}\leq
       C(\|V\|_{L^{1}_{2}})\cdot \langle y\rangle^{2}\cdot
       \|\widehat F_{4}(\cdot,y)\|_{L^{\infty}}
\end{equation*}
and the remaining term
$\|\widehat F_{4}(\cdot,y)\|_{L^{\infty}}$
has already been estimated in \eqref{eq.F4}. Summing up
we have proved that
\begin{equation*}
    |I_{1}|\leq
       C(\|V\|_{L^{1}_{2}})\cdot
       \left(\|g\|_{L^\infty}+\|\HH g\|_{L^{\infty}}\right).
\end{equation*}
The estimate of $I_{2}$ is completely analogous;
the estimate of the terms $II_{1}$ and $II_{2}$ is
even easier, keeping into account that the functions
$m_{\pm}(0,x)$ are bounded on $\R$ (see \eqref{eq.bdd}).

Consider now the more delicate terms $III_{1}, III_{2}$.
Since $m_{-}(0,x)=c_{0}\cdot m_{+}(0,x)$ we can put
the two integrals back together as follows:
\begin{equation*}
    III_{1}+III_{2}=c_{0}
      \int_{0}^{+\infty}d\lambda\int dy
    \frac{m_+(0,y)m_{+}(0,x)}{W(-\lambda)}V(y)
    e^{i\lambda|y-x|}
    \lambda\chi(\lambda)\Im R_{0}(\lambda^{2}+i0)g(y).
\end{equation*}
We decompose this integral in a different way:
\begin{equation*}
    III_{1}+III_{2}=c_{0}IV_{1}+c_{0}IV_{2}
\end{equation*}
where
\begin{equation}\label{eq.IV1}
    IV_{1}=
      \int_{0}^{+\infty}d\lambda\int dy
    \frac{m_+(0,y)m_{+}(0,x)}{W(-\lambda)}V(y)
    \bigl[e^{i\lambda|y-x|}-e^{i\lambda|x|}\bigr]
    \lambda\chi(\lambda)\Im R_{0}(\lambda^{2}+i0)g(y)
\end{equation}
and
\begin{equation}\label{eq.IV2}
    IV_{2}=
      \int_{0}^{+\infty}d\lambda\int dy
    \frac{m_+(0,y)m_{+}(0,x)}{W(-\lambda)}V(y)
    e^{i\lambda|x|}
    \lambda\chi(\lambda)\Im R_{0}(\lambda^{2}+i0)g(y).
\end{equation}
Using the identity
\begin{equation*}
    e^{i\lambda|y-x|}-e^{i\lambda|x|}=\int_{0}^{1}
         e^{i\lambda(s|x-y|+(1-s)|x|)}ds
         \cdot i\lambda\cdot(|x-y|-|x|)
\end{equation*}
and Fubini's theorem we can rewrite $IV_{1}$ as follows:
\begin{equation*}
    IV_{1}=
    \int_{0}^{1}ds\int
    \mathcal F_{\lambda\to\xi}
    \left.\left(
    \frac{\lambda}
         {W(-\lambda)}\lambda\chi(\lambda)
          \uno_{(0,+\infty)}(\lambda)
          \Im R_{0}(\lambda^{2}+i0)g(y)
      \right)\right\vert
       _{\xi=s|x-y|+(1-s)|x|}K\,dy
\end{equation*}
where
\begin{equation*}
    K=K(x,y)=im_{+}(0,y)m_{+}(0,x)V(y)\left(|x-y|-|x|\right);
\end{equation*}
notice that
\begin{equation}\label{eq.Kk}
    |K(x,y)|\leq C|y|\cdot|V(y)|
\end{equation}
by \eqref{eq.bdd}. At this point, we can proceed as above
using Lemma \ref{lem.W} and \eqref{eq.F4} to obtain
\begin{equation*}
    \left\|
    \mathcal F_{\lambda\to\xi}
    \left(
    \frac{\lambda}
         {W(-\lambda)}\lambda\chi(\lambda)
          \uno_{(0,+\infty)}(\lambda)
          \Im R_{0}(\lambda^{2}+i0)g(y)
      \right)
    \right\|_{L^{\infty}_{\xi}}\leq
    C(\|V\|_{L^{1}_{2}})\cdot
           \left(\|g\|_{L^\infty}+\|\HH g\|_{L^{\infty}}\right).
\end{equation*}
whence the estimate of $IV_{1}$ follows immediately.

To conclude the proof, it remains to estimate the term $IV_{2}$.
By property \eqref{eq.media} we have trivially
\begin{equation}\label{eq.zero}
       \int_{0}^{+\infty}d\lambda\int dy
    \frac{m_+(0,y)m_{+}(0,x)}{W(-\lambda)}V(y)
    e^{i\lambda|x|}
    \lambda\chi(\lambda)\Im R_{0}(\lambda^{2}+i0)g(0)\equiv
    0
\end{equation}
(indeed, in the inner integral only $V(y)$ and $m_{+}(0,y)$
depend on $y$). Thus we can subtract \eqref{eq.zero} from
$IV_{2}$ and rewrite it in the form
\begin{equation*}
    IV_{2}=
      \int_{0}^{+\infty}d\lambda\int dy
    \frac{m_+(0,y)m_{+}(0,x)}{W(-\lambda)}V(y)
    e^{i\lambda|x|}
    \lambda\chi(\lambda)
      \bigl[\Im R_{0}(\lambda^{2}+i0)g(y)-
         \Im R_{0}(\lambda^{2}+i0)g(0)\bigr].
\end{equation*}
We now use the elementary identity
\begin{equation*}
    \Im R_{0}(\lambda^{2}+i0)g(y)-
         \Im R_{0}(\lambda^{2}+i0)g(0)=
      \int_{0}^{y}
      \partial_{s}
      \left(\Im R_{0}(\lambda^{2}+i0)g(s)\right)
      ds
\end{equation*}
and we obtain, after applying Fubini's theorem,
\begin{equation}\label{eq.IV22}
    IV_{2}=\int dy
       \int_{0}^{y}ds\,
       m_+(0,y)m_{+}(0,x)V(y)
      \int_{0}^{+\infty}
       e^{i\lambda|x|}
       \frac{\lambda\chi(\lambda)}{W(-\lambda)}
             \partial_{s}
      \left(\Im R_{0}(\lambda^{2}+i0)g(s)
                \right)d\lambda
\end{equation}
Now we notice the following property: by the spectral theorem
\begin{equation*}
    \int_{0}^{+\infty}\phi(\lambda)\lambda\,
                 \partial_{s}
      \left(\Im R_{0}(\lambda^{2}+i0)g(s)
                \right)d\lambda=
    \partial_{s}\phi(\sqrt{H_{0}})g(s),
\end{equation*}
but we have also, using the representation of $\phi(\sqrt{H_{0}})$
as a Fourier multiplier (forgetting constants)
\begin{align*}
    \partial_{s}\phi(\sqrt{H_{0}})g(s)=&\,
         \partial_{s}\int_{\R}e^{i\xi s}\phi(|\xi|)\hat g(\xi)d\xi=
         \int_{\R}i\xi\,e^{i\xi s}\phi(|\xi|)\hat g(\xi)d\xi=\\
         =&
          \int_{\R}
               \,e^{i\xi s}|\xi|\phi(|\xi|)\frac{i\xi}{|\xi|}
                \hat g(\xi)d\xi
\end{align*}
and hence
\begin{equation*}
    \partial_{s}\phi(\sqrt{H_{0}})g(s)=
    C\cdot \sqrt{H_{0}}\, \phi(\sqrt{H_{0}})\,\HH g(s).
\end{equation*}
or, equivalently,
\begin{equation}\label{eq.hil}
    \int_{0}^{+\infty}\phi(\lambda)\lambda\,
                 \partial_{s}
      \left(\Im R_{0}(\lambda^{2}+i0)g(s)
                \right)d\lambda=C
    \int_{0}^{+\infty}\phi(\lambda)\lambda^{2}\,
        \Im R_{0}(\lambda^{2}+i0)\HH g(s)
               d\lambda.
\end{equation}
Thus, by \eqref{eq.IV22} and \eqref{eq.hil}, we obtain
(apart from a constant)
\begin{equation*}
    IV_{2}=
       \int dy
       \int_{0}^{y}ds\,
       m_+(0,y)m_{+}(0,x)V(y)
      \int_{0}^{+\infty}
       e^{i\lambda|x|}
       \frac{\lambda}{W(-\lambda)}
      \lambda\chi(\lambda)\Im R_{0}(\lambda^{2}+i0)\HH g(s)
                d\lambda
\end{equation*}
and this can be estimated exactly as the other terms considered
above.

The proof is concluded.
\end{proof}

\bibliographystyle{plain}

\end{document}